\documentclass[useAMS,usegraphicx]{mn2e}

\title[Lyman Alpha Emission]{Extended and Filamentary Ly$\alpha$ Emission from the Formation of a  Protogalactic Halo at z=2.63 \thanks{This paper includes data gathered with the 6.5 meter Magellan Telescopes located at Las Campanas Observatory, Chile.}\thanks{Some of the data presented herein were obtained at the W.M. Keck Observatory, which is operated as a scientific partnership among the California Institute of Technology, the University of California and the National Aeronautics and Space Administration. The Observatory was made possible by the generous financial support of the W.M. Keck Foundation. }}
\author[Michael Rauch et al.]{Michael Rauch,$^{1}$, George D. Becker,$^{2}$, Martin G. Haehnelt,$^{2,3}$
\newauthor Jean-Rene Gauthier$^{4}$,Wallace L.W. Sargent$^{4}$ \\
$^{1}$Carnegie Observatories, 813 Santa Barbara Street, Pasadena, CA 91101, USA\\
$^{2}$Institute of Astronomy and Kavli Institute for Cosmology, Cambridge University, Madingley Road,  Cambridge CB30HA, UK\\
$^{3}$Kavli Institute for Theoretical Physics, Kohn Hall, University of California, Santa Barbara,  CA 93106-4030, USA\\
$^{4}$Palomar Observatory, California Institute of Technology, Pasadena, CA 91125, USA}

\begin{document}


\pagerange{\pageref{firstpage}--\pageref{lastpage}} \pubyear{2011}

\maketitle


\label{firstpage}

\begin{abstract}  We report the observation of a further asymmetric, extended  Lyman $\alpha$ emitting halo at z=2.63, from our ultra-deep, long-slit spectroscopic survey of faint high redshift emitters, undertaken with Magellan LDSS3 in the GOODS-S field. 
The Ly$\alpha$ emission, detected over more than 30 kpc,  is spatially coincident with a concentration of galaxies visible
in deep broad-band imaging.  While these faint galaxies without spectroscopic redshifts cannot with certainty be associated with one another or
with the Ly$\alpha$ emission, there are a number of compelling reasons why they
very probably form a Milky Way halo-mass group at the Ly$\alpha$ redshift.
A filamentary structure, possibly consisting of Ly$\alpha$ emission at very high equivalent width, 
and evidence for disturbed stellar populations,  suggest that 
the properties of the emitting region  reflect ongoing 
galaxy assembly, with recent galaxy mergers and star formation occurring in the group. 
Hence, the Ly$\alpha$ provides unique insights into what is probably a key mode of galaxy
formation at high redshifts.
The Ly$\alpha$ emission may be powered by cooling radiation or spatially extended star
formation in the halo, but is unlikely to be fluorescence driven by either an AGN or one of the galaxies.
The spatial profile of the emission is conspicuously different from that of typical
Ly$\alpha$ emitters or Lyman break galaxies, which is consistent with the picture that extended emission of this kind
represents a different stage in the galaxy formation process.  Faint, extended Ly$\alpha$ emitters such
as these may be lower-mass analogues of the brighter Ly$\alpha$ blobs.
Our observations provide further, circumstantial evidence that galaxy mergers may promote the production and / or escape of ionizing radiation, and that the
halos of interacting galaxies may be significant sources for ionizing photons during the epoch of reionization.

 \end{abstract}


\section[]{Introduction}

Present-day galaxies are thought to have grown through a combination of mergers  and the continued accretion of gas from
the intergalactic medium. Theoretically, this picture is well motivated (White \& Rees 1978), but  considerable
uncertainties remain about the relative roles of mergers, the inflow of fresh and recycled gas, and the expulsion of gas due to stellar and AGN feedback,
and through interactions.
Observational studies of the {\it emission} from high redshift galaxies have  mostly  concerned themselves with the compact stellar populations of galaxies, whereas narrow-band and spectroscopic studies aiming at the gaseous component have been restricted to the bright end of the luminosity function and the progenitors of the most massive present-day galaxies. Studies of the much less massive building blocks of more typical low-z objects
until very recently had to rely on observations of intervening absorption line systems in the spectra of high redshift background QSOs.
  
At redshift $\sim 3$, a future Milky Way galaxy is expected to consist of a group of mostly faint protogalaxies, spread out over a region spanning several hundred kpc (proper).  QSO metal absorption systems and damped Ly$\alpha$ systems have provided insights into the gaseous environment of these newly forming galactic halos. A picture
of multiple, gaseous protogalactic clumps, accreting gas and ultimately merging, has been used to successfully model the properties of observed QSO absorption systems (e.g., Haehnelt, Steinmetz,\& Rauch et al 1996, Rauch, Haehnelt \& Steinmetz et al 1997). 
Theoretical modeling  has also shed light on the likely geometry of the gas accretion process and the relative importance of the various accretion modes
(e.g., Birnboim \& Dekel 2003; Keres et al 2005; Dekel \& Birnboim 2006, van de Voort et al 2011; Fumagalli et al 2011). 

Ultra-deep spectroscopic surveys (e.g., Bunker et al 1998; Santos et al 2004; Rauch et al 2008, 2011) have reached surface brightness limits where it becomes possible to trace the progenitors of present-day galaxies and some of their protogalactic building blocks   directly in Ly$\alpha$ emission, at radii
out to tens of kpc. The deep spectroscopic FORS-VLT survey by Rauch et al (2008; see also Barnes \& Haehnelt 2010; Rauch \& Haehnelt 2011) 
appears to have detected the bulk of the host galaxies of damped Ly$\alpha$ systems (DLAS) and Lyman limit systems (LLS) at $z\sim3$
in the form of faint Ly$\alpha$ emitting galaxies, with space densities reaching $3\times 10^{-2}$ h$_{70}^3$ Mpc$^{-3}$.   Clustering studies of the generally brighter Lyman $\alpha$ emitters found by narrow-band imaging surveys suggest that these studies  are probing the massive end of the same progenitor population  (e.g., Guaita et al 2010). 

\medskip 

\begin{figure*}
\includegraphics[scale=.55,angle=0,keepaspectratio = true]{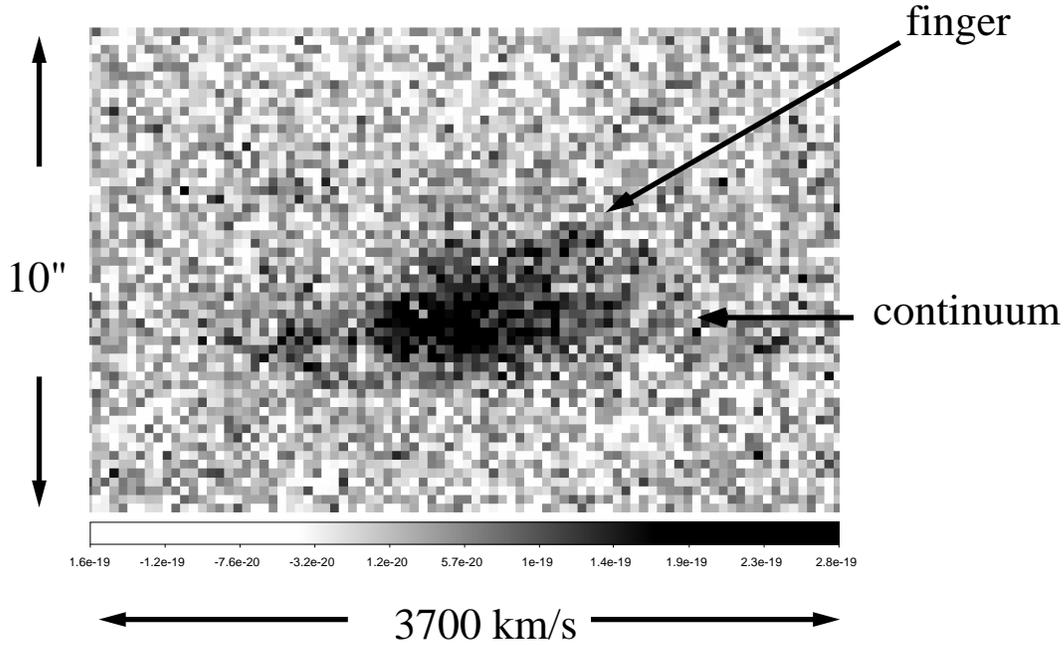}
\caption{Spectrum of the z=2.63 Ly$\alpha$ emission line. The wavelength increases toward the right, the vertical direction is the spatial coordinate along the slit, which was oriented at a position angle of
0$\ \deg$, i.e., N-S. The spatial extent along the slit (N-S) of the section shown here is 10", the velocity extent in the spectral (horizontal direction) is about 3700 kms$^{-1}$. 
The 1-$\sigma$ fluctuation in the flux density per $0.185"\times0.64\AA $ wide pixel is $5 \times 10^{-20}$erg cm$^{-2}$s$^{-1}$ $\AA^{-1}$. The arrow shows the approximate horizontal position of
the strongest continuum trace (caused by object A, see fig. \ref{prettypics}), and a spatially extended Ly$\alpha$ "finger". \label{specraw}}
\end{figure*}

The progenitors of present-day galaxies are expected to undergo repeated mergers. Interactions of the merging constituents with the ambient halo medium  and among  each other may transform the properties of the stellar populations
and the physical state of the gas in ways observable even at high redshift.  For example, intriguing evidence for the connection between Ly$\alpha$ emission and the small scale galactic environment has been presented by Cooke et al (2010), who found
spatially close pairs of Lyman break galaxies to show a very high fraction of  Ly$\alpha$ emitters. The Ly$\alpha$ photons may be escaping because
of damage to the optically thick gaseous halo during encounters  (e.g., Hibbard, Vacca \& Yun 2000), or may result from star-formation, perhaps induced in the halo by tidal tails or ram-pressure stripped gas, or by nuclear inflows. These effects may also work to increase the production and favor the release of ionizing radiation, turning mergers into promising sites for the sources of ionizing photons during the epoch of reionization 
(e.g., Bridge et al 2010; Rauch et al 2011, hereafter paper I). 

Immediate observational signatures of Ly$\alpha$ emitters caused by interactions  may include spatially extended Ly$\alpha$ emission, a disturbed velocity field imprinted on the Ly$\alpha$ emission line, and the presence of multiple sources of continuum and ionizing radiation. These characteristics suggest that these objects are best approached by a combination of deep spectroscopy (for the detection and velocity structure of the gaseous halos) with deep space-based imaging (to study the stellar populations  and the geometry).

Two such surveys have now been performed by us in the Hubble Deep Field North (HDFN, with Keck I / LRIS)  and
and the Hubble Deep Field South (HUDF, with Magellan/LDSS3), and the respective GOODS flanking fields, taking advantage of existing  space-based imaging. Among the generally compact Ly$\alpha$ emitters, a small fraction are found that show pronounced spatial asymmetries relative to a continuum
source, and extended morphologies (diameters of several arcseconds, to the detection levels reached) in Ly$\alpha$ emission. With a space density of $\sim 10^{-3}h_{70}^{3}$ Mpc$^{-3}$, and fluxes of a few times $10^{-17}$erg\ cm$^{-2}$ s$^{-1}$
these objects are more common and fainter than the so-called "Ly$\alpha$ blobs" (e.g., Francis et al 1996; Fynbo, M\o ller, \& Warren 1999;  Keel et al 1999; Steidel et al 2000; Palunas et al 2004; Matsuda et al 2004, 2011, 2012; Dey et al 2005;  Nilsson et al 2006; Ouchi et al 2009; Yang et al 2009, 2010, 2011; Scarlata et al 2009, Prescott et al 2009, 2012, Bridge et al 2012), although, as we shall discuss below, they share some similarities.

Of the three brightest objects matching these criteria in the HUDF, the first one is the interacting  $V\sim27$, z=3.34 object described in paper I, the second is the  object to be described here, and the remaining one (to be published) turned out to be the optical counterpart of an obscured x-ray bright z=3 QSO.

The present paper is organized as follows:  the observations of the z=2.63 Ly$\alpha$ halo are presented, starting with an analysis of the spectroscopic properties of the Ly$\alpha$ emission
line region. We then examine HST ACS images for clues about the underlying stellar populations. The detection of a filamentary  structure
likely to be glowing in Ly$\alpha$ is described in the next section, followed by a discussion of possible origins of the radiation. The nature of the underlying association of objects
is examined in the context of dark matter halos, concluding with a comparison between the extended emitters discussed here and the more common "compact" Ly$\alpha$ emitters 
and Ly$\alpha$ emitting Lyman break galaxies. Simple models for the surface brightness distribution of the compact emitters are used to get insights into the nature
of the underlying gaseous halo. An appendix consisting of three sections describes (A) a technical point about how to determine the actual position of the spectrograph slit retrospectively
from a comparison of projected spectra and imaging; (B) how a lag between an instantaneous starburst and Ly$\alpha$ recombination radiation can maintain detectable Ly$\alpha$ emission and a high equivalent width beyond the decline of the stellar sources of ionizing radiation; and (C) how the  observations of the high S/N  surface brightness profiles of individual Ly$\alpha$ emitters
from our surveys can constrain the line formation process.

\section[]{Observations}

The object described here was discovered in the long-slit, spectroscopic, blind survey in the Hubble Ultra Deep Field (HUDF)/GOODS-South field (Beckwith et al 2006; Giavalisco et al 2004) described in paper I. The data are based on a spectrum taken with the LDSS3 spectrograph
on the Magellan II telescope at Las Campanas. Using a $2"\times8.3'$ long slit at a position angle of 0 degrees and the VPH blue grating, a set of 3000s long exposures
totaling 61.4 hours of useful exposure time were recorded during November 18-23, 2008 and November 11-16, 2009.
The spectra consist of 0.189" wide pixels, with a slit-width limited resolution of 340 kms$^{-1}$ (FWHM). The useable survey volume was $2056h_{70}^{-1}$ Mpc$^{-3}$,
at mean redshift 3.33.

\subsection[]{Spectroscopic data}

The object is the second of three spatially extended, asymmetric (with respect to the brightest spectroscopically detectable continuum underlying the 
Ly$\alpha$ emission),  relatively bright (F$>2 \times 10^{-17}$erg cm$^{-2}$s$^{-1}$) Ly$\alpha$ emitters in the survey. 
The Ly$\alpha$ line peaks at 4414.4\AA\ (fig.\ref{specraw}), corresponding to redshift
z=2.631. 
 The total line flux observed through the 2" wide slit amounts to $(3.7\pm 0.2) \times 10^{-17}$erg cm$^{-2}$s$^{-1}$ (statistical errors only), which must be considered a lower
limit, given the possibly considerable aperture losses. Similarly, the observed equivalent width, found  to be 31.7A in the rest frame, from a comparison
of the Ly$\alpha$ flux to the flux the brightest continuum trace immediately redward of the Ly$\alpha$ line, is a measure of limited usefulness because of its aperture dependence.
The maximum spatial extent of the Ly$\alpha$ emission line in the vertical direction
(i.e., N-S) along the slit is about 3.9" (32 kpc proper).  In the spectral direction the flux density is detectable at the $\sim 10^{-19}$ erg cm$^{-2}$ s$^{-1}$ $\AA^{-1}$ level over about 1800 km s$^{-1}$.
The line consists of a clumpy, multi-humped profile with a dominant asymmetric red peak, showing an extended red shoulder. The width of the red peak if crudely fitted by a Gaussian profile
is about 870 km s$^{-1}$ at the location of the brightest continuum trace (see arrow in fig. \ref{specraw}).

\begin{figure*}
\includegraphics[scale=.65,angle=0,keepaspectratio = true]{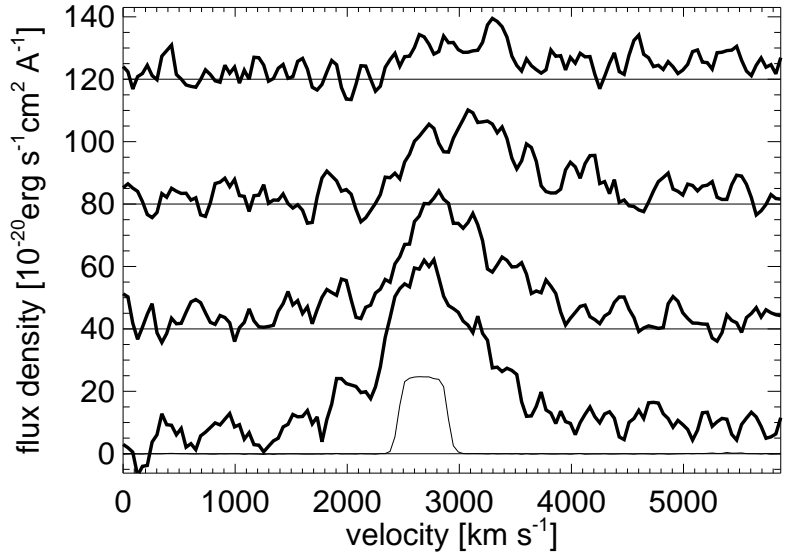}
\caption{1-dimensional spectra, representing 4 horizontal cuts through the spectrum in fig. \ref{specraw} (with their N-S positions indicated by the white horizontal lines straddling
the left and center panel of fig. \ref{prettypics} below). The first three cuts are offset by 0.75" N with respect to each other (the northernmost one offset such as to line up with the Ly$\alpha$ finger extending NE from the main clump). The bottom spectrum in the current figure is the southernmost cut and roughly lines up with the continuum produced by objects A and C (see fig. \ref{prettypics}, for the nomenclature). The second spectrum from the bottom in the current figure approximately runs through the northernmost part of the elongated structure "T2", the third spectrum from the bottom
through the object "B", and the topmost one through "T3". The small peak in the topmost spectrum corresponds to the 
finger of Ly$\alpha$ emission in fig. \ref{specraw}. The spectra are produced from the two-d spectrum by smoothing with a 1.75 pixel wide Gaussian kernel and extracting
a 3 pixel-wide box profile. The flux density ordinate refers to the flux density gathered in a 3-pixel wide spectrum. The spectra have been offset by identical amounts along the flux density ordinate
for clarity. The thin boxy profile at the bottom is the measured line profile for unresolved emission filling the slit (from a HeNeAr calibration lamp).\label{onedspec}}
\end{figure*}

\begin{figure*}
\includegraphics[scale=.65,angle=0,keepaspectratio = true]{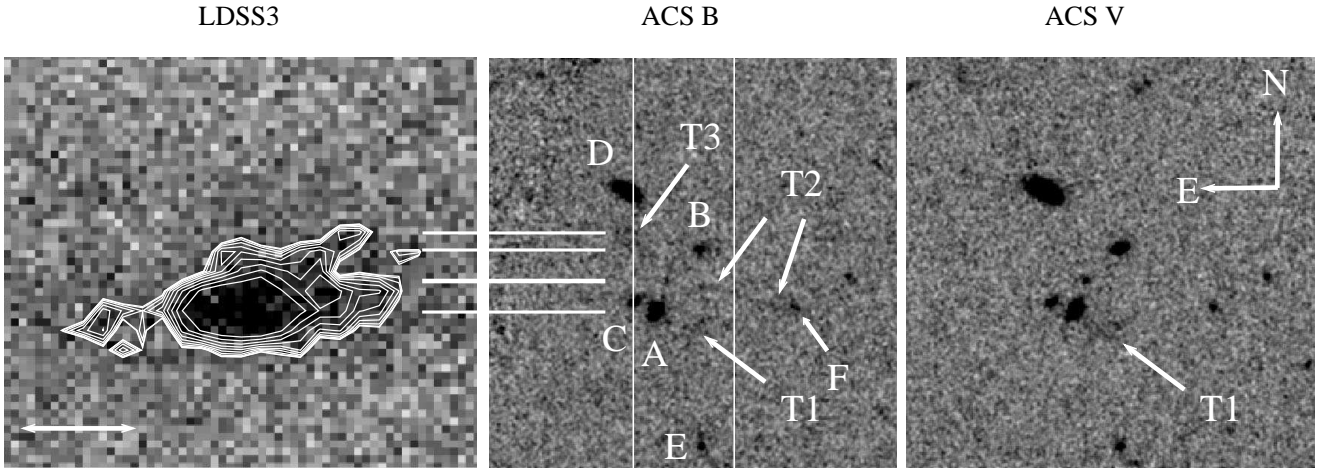}
\caption{Ly$\alpha$ line spectrum and GOODS-South  ACS imaging. Left panel: The LDSS3 spectrum of the Ly$\alpha$ line at 4414 \AA . For better visibility of the lower
light level region we have
added white 
contours showing the Ly$\alpha$ emission flux density. The outermost contour
corresponds approximately to $10^{-19}$ erg cm$^{-2}$ s$^{-1}$ $\AA^{-1}$. The length of the white double arrow indicates a velocity scale of 600 kms$^{-1}$. Middle Panel:
10"x10" B band ACS image, centered on 03:32:38.821 -27:44:27.90 (2000). The two solid vertical lines denote the position of the slit edges, as determined with the method described in Appendix A. Object D is a z=0.663 foreground object. Several of the other galaxies may be at the redshift of the Ly$\alpha$ halo. An extended stellar, possibly tidal patch of emission is indicated by the arrow marked with "T1". A thin filament of emission ("T2") appears to connect another faint compact object ("F") with what appears to be an even fainter one close to the center of the field. A patchy spot of emission ("T3") can be discerned
just south of the object D. The horizontal white lines indicate the positions along the slit in S-N direction  of the four one-dimensional spectra in fig. \ref{onedspec}. Right Panel: ACS V band image.  
Note the persistence of feature T1 but not T2 or T3, as compared to the B band image. \label{prettypics}}
\end{figure*}

Fig.\ref{onedspec} shows four one-dimensional spectral slices, with the first three offset by 0.75"  with respect to each other in the direction along the slit (i.e., S-N; the last one has a slightly smaller offset made to line up with an extended "finger" of Ly$\alpha$ emission).  The positions of the spectra along the slit are also indicated in fig.\ref{prettypics} by the four horizontal lines.

As seen in fig. \ref{onedspec}, the width of the red peak initially appears to change little, going South to  North, at least between
the first (southernmost) and third position. However, the centers-of-flux in the red peak shift by approximately 310 km s$^{-1}$ between the 
first and  the third profile. The fourth and last spectrum, positioned spatially to intersect the finger of Ly$\alpha$ emission that extends outward from objects A and C in a northern direction (see fig. \ref{prettypics}), shows a 
now much narrower peak (FWHM$\sim  190$ km s$^{-1}$; see also the 2-dimensional spectrum in figs.\ref{specraw} and \ref{prettypics}) that is shifted further to the red from the previous cut by another 250 km s$^{-1}$.

The interpretation of the line widths depends crucially on the position of the emitting gas in the direction across the slit.
If the Ly$\alpha$ emission is sufficiently spatially extended to fill the slit, then different positions in the direction across the slit amount to  noticeable shifts in observed wavelength on the detector. In this case, 1" W corresponds to 3.32 \AA\ to the red. Thus, the E-W separation between the objects A/C and B alone would shift any  Ly$\alpha$ emission from these objects
by 250 km s$^{-1}$, even if there were no velocity difference between the two locations.  The full velocity extent across the slit amounts to 460 km s$^{-1}$ FWHM. The actual spectral response 
for an unresolved, spatially extended emission line is shown as the boxy, thin solid line in fig. \ref{onedspec}.

\begin{table*}
\scriptsize
 \centering
 \begin{minipage}{170mm}
  \caption{GOODS-S  objects}
  \begin{tabular}{@{}rllcccccc}
\hline 
 ID  & z$_{\rm spec}$ & z$_{\rm phot}$& GOODS$^c$ & COMBO-17$^d$/Music$^e$ & V (F606W)$^c$  &B-V  \\
 \hline
A & 2.631$^{a}$, 2.6191$^{b}$,2.619$^{f}$& 1.91$^d$, 1.49$^e$, 2.066$^f$ & J033238.89-274429.1    & 40596 / 13597 &  25.12$\pm$0.02$^c$&0.20$\pm$0.03$^c$&   \\
B & 2.631$^{a}$(?) & 0.043$^d$, 3.06$^e$ & J033238.81-274427.6    & 40626 / 13635 &  26.06$\pm$0.04$^c$&1.01$\pm$0.11$^c$  &   \\
C & 2.631$^{a}$(?) & --- & J033238.93-274428.9    & ---  / --- &  27.33$\pm$0.06$^c$& 0.25$\pm$0.11$^c$ &   \\
D & 0.663$^{a}$ & 0.681$^d$, 0.69$^e$ & J033238.94-274426.2    & 40690 / 13640 &  24.46$\pm$0.02$^c$& 0.76$\pm$0.04$^c$ &   \\
E & ? & 2.092$^f$ & J033238.80-274432.4    & --- / --- &  27.92$\pm$0.08$^c$&0.34$\pm$0.15$^c$  &   \\
F & ? &    & J033238.63-274429.0    & ---  / --- &  28.92$\pm$0.18$^c$& 0.27$\pm$0.32$^c$ &   \\
filament T2 & ? &$2.2 < z< 2.9^h$    & ---   & ---  / --- &  $> 29.44$  & $<-1.71^g$ &   \\
\hline
\end{tabular}
comments: a) this paper; b) Balestra et al 2010; c) Giavalisco et al. 2004; from the h\_goods\_sv\_r1.1z.cat.txt catalog file; d) Wolf et al 2001; e) Grazian et al 2006; f) Cardamone et al 2010; g) $1\sigma$; in custom aperture.; h) from the wavelength limits of the B band filter which would limit the redshift of the presumed Ly$\alpha$ emission to those values.
\end{minipage}
\end{table*}

\subsection[]{Broad band detections of galactic counterparts}

The position of the Ly$\alpha$ emitter along the slit coincides in projection with what appears to be a small group of galaxies within a few arc seconds of each other (fig. \ref{prettypics}, middle and right panels). The position of a continuum trace (hard to see with the color stretch applied here, and  
indicated in fig. \ref{specraw}  by a horizontal arrow)  appears offset to the south from the peak of the extended emission. 
It is caused mainly by an object at 03:32:38.888 -27:44:29.07 we denote as "A" in fig.\ref{prettypics}, with a fainter nearby object here called "C".  The Ly$\alpha$ emission appears to stretch North (and redward) in a "finger" of emission reaching somewhat beyond the declination of object "B", suggesting that B may possibly belong to the same halo as A,  and so may C.  The object "D" is a foreground interloper. Its redshift can be identified in our spectrum from an [OII] 3728 doublet as z=0.663. The centroids of the continuum traces of objects D and A are separated by 2.8" along the slit (i.e., in the North-South direction). The names in the literature for the objects and the various redshift measures are given in table 1.

There are several strands of evidence suggesting that several objects shown in fig. \ref{prettypics} are part of the same galactic halo,
which we may be observing during the infall of multiple protogalaxies.
At least two other objects occur within less than about 6" (48.8 kpc proper) 
of A, with similar B-V colors, namely  "E" and "F". Moreover, two extended regions of emission  with an appearance reminiscent of tidal
features ("T1" and "T2") are visible to the West and Southwest of A, and a faint extended region T3 is seen just South of D and East of B.
Both B and T3 in the image are lining up in projection with the narrow Ly$\alpha$ "finger" extending to the upper right in the 2-d spectrum (e.g., fig. \ref{specraw}). Below we shall argue that at least T3 may be linked causally to the Ly$\alpha$ "finger".

The faint region "T1" extends about 1.3"
to the SW of object A and appears to consist of continuum sources, as it is visible in both B and V bands.
Additionally, in the B band image, a thin filament of emission ("T2") appears to extend 3.5" West from galaxy A, ending in another very faint object, "F".
The T2 structure, which we shall discuss further below, is invisible in the V band, as is T3.

\subsubsection[]{The puzzle of the discrepant photometric redshifts}

We note that the photometric redshifts for objects A and B from the COMBO-17 (Wolf et al 2001),  GOODS-CFDS-Music (Grazian et al 2006) and MUSYC (syc!)  (Cardamone et al 2010) collaborations do not agree well with each other or with the spectroscopic Ly$\alpha$ redshifts  (table 1). The difficulty of separating spatially close objects in the partly ground-based data, their faintness, and the presence of Ly$\alpha$ emission in the blue filter 
may be partly to blame here. However, given our rather peculiar selection criterion
(extended, asymmetric Ly$\alpha$ emission) it is conceivable that the underlying
objects are also peculiar, in the sense that their broad-band colors may not have been anticipated by the usual photometric redshift techniques. Indeed, object A is unusually bright in
the U and B bands for z=2.63 which must have led the surveys
to assign a lower redshift than indicated spectroscopically by the Ly$\alpha$
line.  

\begin{figure*}
\includegraphics[scale=.55,angle=0,keepaspectratio = true]{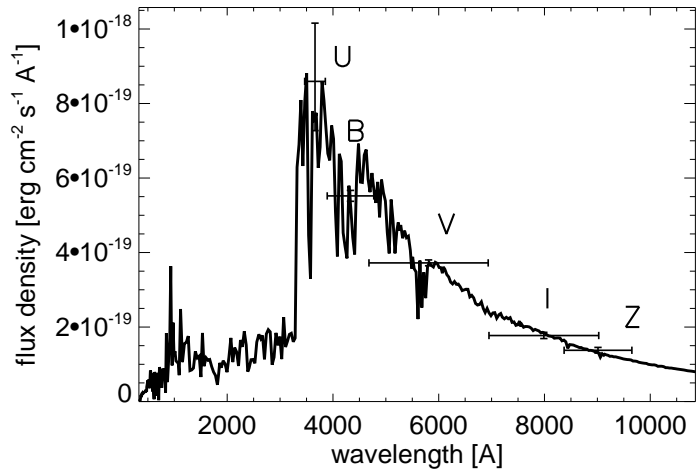}
\caption{Best fit of the broad band measurements for galaxy A with filters WFI U38, and HST ACS B,V,I, and Z from Grazian et al (2006) using a Starburst99 model spectrum (Leitherer et al. 1999), with total stellar mass $6.7\times10^7$ $M_\odot$
and age $4\times10^6$ yr, redshifted to z=2.63 and attenuated by the average Ly$\alpha$ forest opacity blueward
of 1215.67 \AA\ in the rest frame.  \label{starburst}}
\end{figure*}

Object "A" may host an AGN, which, at first sight, may seem consistent with the relatively large line width (870 kms$^{-1}$) of the
emission line.
However, the fact that the Ly$\alpha$ emission line width is spatially extended, retaining its spectral line width  over several arcseconds in the North-South
direction away from the continuum trace of A suggests that this is not the broad emission line of an AGN, which would be
spatially compact. Moreover, we have argued above that the line width may be partly caused by the spatially extended emission and the wide slit. 
There are currently no other signs (high ionization transitions, radio- or x-ray detections) to support
an AGN interpretation. The HeII 1640 region is somewhat damaged in the spectrum and does not allow us to make a meaningful measurement.
We note that there is a nearby object ("E", in our table) for which the MUSYC survey (Cardamone et al 2010) quotes a very similar photometric redshift ($z_{phot}=2.092$) as for object "A" ($z_{phot}=2.066$). The most likely explanation is that A is at the spectroscopic redshift of Ly$\alpha$ (indicated also by a faintly visible Ly$\alpha$ forest decrement in its continuum); that A and E with their similar colors are indeed at the same redshift, and that both photometric redshifts are off by similar amounts. It is conceivable that these are physically similar objects made unusually blue by the same mechanism. Both may have undergone recent starbursts, which may have produced  
objects brighter in the U- and B-bands than the templates used for photometric redshift determinations. We have fitted the GOODS-CFDS-MUSIC broad band magnitudes of object A in the ESO WFI U38 (hereafter "U"), 
HST ACS WFC F435W ("B"), F606W ("V"), F775W ("I"), and F850LP ("z") bands with
spectral synthesis models of instantaneous starbursts and continuous star-formation
included in the Starburst99 program (Leitherer et al 1999). Good fits at the observed Ly$\alpha$ redshift were obtained
for a burst with the default input parameters, plus a total stellar mass of $6.7\times10^7 M_\odot$, a metallicity of 0.020 solar, and an age
of $4\times10^6$ yr (fig. \ref{starburst}). Continuous star formation with a star formation rate of 14 $M\odot$yr$^{-1}$ observed at a very similar age
gave a similarly good representation of the bluer band colors. The rest frame slope of the UV continuum $\beta$, with $f_\lambda\propto \lambda^{\beta}$ determined
from the $V-I$ and $U-V$ colors, is
$\beta(V-I)=-2.3161$,  and $\beta(U-V)=-2.046$, respectively. These continuum colors are rather blue  compared to those of known Ly$\alpha$ emitters  at similar redshifts (e.g., Venemans et al 2005), and presumably indicate the absence of dust. With a $\beta<-2$ the object appears similar
to the less evolved star-forming objects observed at redshifts possibly associated with the epoch of reionization (e.g.,  Bouwens et al 2010; Finkelstein et al 2010; Dunlop et al 2012).

The formation of such a young stellar population  may have been triggered by the interactions between the members
of the protogalactic group. Gas dynamical disturbances and rapid star formation would naturally lead to enhanced and messy Ly$\alpha$ emission, with obvious
similarities to the earlier case discussed in paper I. If the galaxies entered their future common halo with a pre-existing stellar population the interaction
would lead to a composite population with the young stars superposed on that  older stellar population.  An attempt at extracting the
properties of the older stars by fitting the observed broad band magnitudes with the MAGPHYS SED fitting software (da Cunha et al 2008), using a larger set of filters including the U,B,V,I and z used with
starburst99 plus restframe optical VLT H,J,K,and Spitzer IRAC 3.6, 4.5, and 5.8 $\mu$ data as collated by the GOODS-MUSIC collaboration (Grazian et al 2006), yields a stellar population of $1.1\times10^9 M\odot$ with 
formation age $7\times10^8$yr. Omission of the U and B filters (which are not well fit) produced no changes in the parameters from the MAGPHYS output but shrank the reduced $\chi^2$ from
1.30 to 1.04. Thus the MAGPHYS SED fitting of the rest-frame optical colors is picking up the older population of object A, dating from before the recent star-formation event.

\subsection[]{Broad band detection of a Ly$\alpha$ emitting filament}

\begin{figure*}
\includegraphics[scale=.65,angle=0,keepaspectratio = true]{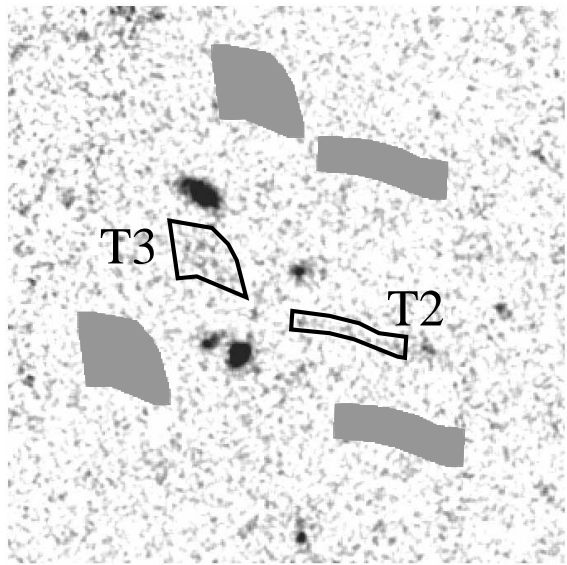}
\caption{ACS WFC B-band image showing the extraction apertures (black) used for the filament T2 and another more amorphous emission region T3, as well  as two background windows (solid) for sky subtraction and noise measurements for each aperture. 
\label{masks}}
\end{figure*}

Further insights into the physical state of the halo are provided by the presence of the faint, filamentary thread of emission (marked "T2") in 
the B band exposure, shown in the central panel of fig. \ref{prettypics}. The thread appears to extend continuously from (03:32:38.800 -27:44:28.5) to (03:32:38.631 -27:44:28.99)
i.e., over 2.3" (18.7 $h_{70}^{-1}$ kpc proper). Its width can be traced over about 4 pixels or 0.12" and so it may be
be spatially unresolved in the short dimension. It terminates at either end in brighter spots, of which at least the western (brighter) one (object F) contains stars as it is
visible in both B and V bands. This filament is only partly covered by our slit, but its position along the slit is consistent with it being
co-spatial with the Lyman alpha emitting region, at least in projection.  Its apparent connection with a galaxy  makes it
unlikely that the filament is an artifact caused by the instrument or by the data reduction.
The structure is remarkable in that it is absent from the (deeper) V band (and not visible in any of the other ACS bands either).
We measured the flux in an irregular aperture drawn around the filament (see fig.\ref{masks}). The sky background and the noise were measured from 
offset background windows placed in apparently clean areas of the image. 
The measured sky-subtracted fluxes in the filamentary aperture in the GOODS ACS mosaic B and V band images amount to 0.151$\pm$0.032 counts (F435W) and -0.0979$\pm$0.0654 counts (F606W), i.e. the filament is a 4.7 $\sigma$ detection in the B band, and a non-detection in the V band.
 
Using the published ACS GOODS zero points of 25.67 (F435W) and 26.49 (F606W) we obtain AB magnitudes for the filamentary aperture
of 27.73 (F435W) and $>29.44$ (1$\sigma$; F606W), or a $1\sigma$ upper limit B-V=-1.71.  Interestingly, the wavelength of the Ly$\alpha$, if at the redshift of the main halo, falls close to the central wavelength
of the B filter (transmission 92\%). Thus the very blue color may be caused by a dominant contribution from Ly$\alpha$ emission to the broad band count rate.  More precisely,
if the flux were indeed Ly$\alpha$, then the width of the B filter would limit the redshift of the filament to be within  $2.2 < z < 2.9$, which is consistent with it
belonging to the same halo as the spectroscopically detected, main Ly$\alpha$ emission.
We can estimate the hypothetical Ly$\alpha$ flux from the B-V difference, assuming that the underlying stellar population is similar to color-selected ("Lyman break") galaxies 
(e.g., Shapley et al 2006) in that its continuum  is essentially flat in $f_{\nu}$. 
Ascribing the flux in the B-band to the sum of line ($f_{\nu}^{ly\alpha}$) and continuum ($f_{\nu}^ {\rm cont}$) emission, and the flux in the V-band (or rather the upper limits) entirely to continuum emission, we can then write 
\begin{eqnarray} 
f_{\nu}^{ly\alpha} = <f_{\nu}^B> - <f_{\nu}^{\rm cont}> \approx  <f_{\nu}^B> - <f_{\nu}^V>,
\end{eqnarray}
where it has been assumed that the flux per frequency interval is flat between B and V.

The resulting total Ly$\alpha$ flux $F_{Ly\alpha}$ responsible for the observed B-V counts can then be estimated from
\begin{eqnarray} 
(<f_{\nu}^B> - <f_{\nu}^V>)=\left(\int{f_{\lambda}^{ly\alpha} \lambda T_{\lambda}^B d\lambda}\right)\left(c\int{\frac{T_{\lambda}^B}{\lambda} d\lambda}\right)^{-1}\nonumber\\ 
\approx  F_{Ly\alpha}\lambda_{ly\alpha} T_{ly\alpha}^B \left(c\int{\frac{T_{\lambda}^B}{\lambda} d\lambda}\right)^{-1}
\end{eqnarray}
 as $F_{Ly\alpha}= (4.0\pm 1.2)\times 10^{-17}$erg cm$^{-2}$s$^{-1}$. Here $T_{\lambda}^B$ is the transmission of the B filter, and the additional
$\lambda$ terms in the integrals account for the fact that the count rate is proportional to $\lambda f_{\lambda}$. 
The total flux from the filament in the B-band is somewhat larger than the total Ly$\alpha$ flux in the spectrum that has passed through the spectrograph slit. However, the filament is mostly outside of the slit,
so the Ly$\alpha$ line in the spectrum is only partly due to the filament, and the contribution of the filamentary emission
may not dominate the spectroscopically observed emission.

\medskip

An assessment of the equivalent width
is very uncertain, because of the asymmetric and non-linear errors as we do not have a detection of the continuum, but we can get an idea
of the uncertainties involved by trying to maximize the uncertainties in a correlated way such as to favor as low an equivalent width as possible.
At the $1\sigma$ level, if we simultaneously subtract a $1\sigma$ deviation from the B band flux and add such a deviation to the V band flux (and instead of the
formally negative flux we conservatively impose a positive V band flux with the size of a $1\sigma$ error), we find that
such fluctuations would result in a rest frame equivalent width lower limit of 688 A (1$\sigma$). If we had used only the
error in V but assumed a $3\sigma$ fluctuation, the allowed rest frame equivalent width would still be 149 \AA . It is clear however, that the data is also consistent with no stellar continuum, and the much higher upper bound on the equivalent width
given by the 2-photon continuum (e.g., Dijkstra 2009). 

\medskip

We briefly point out the presence of another faint patch of emission, shown as "T3" in fig. \ref{prettypics}, that is also visible in the B but not in 
the V band, with a $3\sigma$ upper limit on the B-V color of $< -1.6$. Unlike the filament T2, this structure is not connected to any other object so we have less reason to consider it as  real or as belonging to
the halo, other than it being a 7.9$\sigma$ detection in the aperture shown in fig \ref{masks}. It lines up well, however, in the direction along the slit with the narrow "finger" of Ly$\alpha$ emission seen in the 2-d spectrum (fig.\ref{prettypics}; see the upper most two spectral cuts). If the B-band
light were dominated by Ly$\alpha$ flux, 
the resulting rest frame Ly$\alpha$ equivalent widths indicated by the upper limits on the V-band flux would be between 426 (3$\sigma$) and 1768 \AA ($1\sigma$),
again assuming a flat spectrum. Increasing the contribution
of a hypothetical stellar continuum to the B-band by changing the continuum slope $\beta$(B-V) from flat (=-2) to an extreme value of -4, would 
only increase the B-V by 0.6. If the B-band light for T3 were entirely due to Ly$\alpha$ flux, the flux would be
$(8.3\pm 1.5)\times10^{-17}$ erg cm$^{-2}$s$^{-1}$, whereas the spectroscopically observed fraction of the flux in the region of the 2-d spectra that coincides with the T3 structure in projection is about $1.2\times10^{-17}$ erg cm$^{-2}$s$^{-1}$. T3 straddles one of the slitjaws, so much of
the light may not fall onto the slit.  We estimate crudely that slit losses of 2/3 would correct the total spectroscopically detectable
flux to $3.6\times10^{-17}$ erg cm$^{-2}$s$^{-1}$. This is within about 3$\sigma$ of the flux in the B-band. 
We conclude that it is possible that the apparent B-band flux in the region T3 may be real Ly$\alpha$ emission, partly seen in the spectrum as well.
The Ly$\alpha$ "finger" visible in figs. \ref{specraw} and \ref{prettypics} that appears to correspond to the T3 emission shows up
in the 1-dimensional cuts (fig. \ref{onedspec}, top spectrum) as the little narrow peak near 3400 kms$^{-1}$. The small line width (190 kms$^{-1}$ FWHM)
implies that the source does not fill the slit, and the single, symmetric, narrow line may reflect optically thin (i.e., partly ionized) gas. 

\subsection[]{Sources of the observed Ly$\alpha$ emission}

In this section we are looking at the observed Ly$\alpha$ budget, distinguishing between the total flux observed
spectroscopically throught the slit, and the presumed flux from the extended components T2 and T3 as seen 
in the B-band imaging. 

For the {\it total spectroscopically observed Ly$\alpha$ flux} to be produced by the stellar sources present, the ionizing photons expected from the rest frame 1500\AA\ luminosity $L_{1500}$ of the broad band objects have to be able to provide the photoionization rate  required.
Estimating the stellar yield of ionizing photons as in paper I for the three brightest $z\sim 2.63$ sources A, B, and C, we arrive at $\dot{N}^{{\rm ion}}_{*}=6.6\times10^{53}$s$^{-1}$,
whereas the production rate required by the observed Ly$\alpha$ emission is 
$\dot{N}^{{\rm Ly}\alpha}_{*}=2.0\times10^{53}$s$^{-1}$, about 30\% of the stellar rate, so the number of ionizations needed to account for the
overall Ly$\alpha$ flux that went through the spectrograph slit can be accounted for as photoionization from the stellar sources thought to be associated with the Ly$\alpha$ halo.

\bigskip

The astrophysical source of the {\it filamentary Ly$\alpha$ emission} is not immediately obvious. 
Below we will discuss several possibilities, including fluorescence in response to ionizing radiation from one of the nearby galaxies, or from an obscured AGN; cooling radiation
in a cold accretion filament; shocks and emission from windshells; and intra-halo star formation, including special conditions like more massive or metal-poor stars
favoring large yields of Ly$\alpha$ and ionizing photons.

\subsubsection[]{Fluorescence caused by the loss of ionizing photons from one of the other galaxies in the halo}

The filamentary Ly$\alpha$ may be  caused by fluorescence in
response to ionizing photons from galaxy A.  As this was the preferred explanation for the earlier, Rauch et al 2011 object, we will scrutinize it here in some detail.
The number of ionizations required to occur in the filament,
\begin{eqnarray}
\dot{N}^{{\rm ion}}_{fil}=2.15\times10^{53} s^{-1}\left(\frac{F_{ly\alpha}}{4\times10^{-17}}\right)\left(\frac{1}{1-f^{ll}_{esc}}\right)\left(\frac{1}{f^{ly\alpha}_{esc}}\right),
\end{eqnarray}
if produced by escaping ionizing radiation from galaxy A which is at a projected distance R=16.6 kpc from the center of the filament, would amount to
\begin{eqnarray}
\dot{N}^{{\rm ion}}_{fil}=\left(\frac{3.98\times10^{55}}{s}\right)\left(\frac{F_{ly\alpha}}{4\times10^{-17}}\right)\left(\frac{1}{1-f^{ll}_{esc}}\right)\left(\frac{1}{f^{ly\alpha}_{esc}}\right)\nonumber\\ \times \left(\frac{R}{16.6{\rm kpc}}\right)^2\left(\frac{\Delta A}{18.7{\rm kpc}^2}\right)^{-1}.
\end{eqnarray}
Here it was assumed, that 2/3 of the ionizing photons get converted into Ly$\alpha$, and we have computed the number of ionizations in terms
of the escape fraction for ionizing photons, $f^{ll}_{esc}$, the escape fraction for Ly$\alpha$ photons, $f^{ly\alpha}_{esc}$,  and the area $\Delta A$
subtended by the filament. 
If the ionizing radiation were purely of stellar origin, the necessary luminosity density $L_{1500}$ would be 
\begin{eqnarray}
L_{1500}^A=4.7\times10^{30} {\rm erg\ }{\rm s}^{-1}{\rm Hz}^{-1}\nonumber\\
\times\left(\frac{F_{ly\alpha}}{4\times10^{-17}}\right)\left(\frac{1}{1-f^{ll}_{esc}}\right)\left(\frac{1}{f^{ly\alpha}_{esc}}\right)\times \nonumber\\ \times\left(\frac{R}{16.6{\rm kpc}}\right)^2\left(\frac{\Delta A}{18.7{\rm kpc}^2}\right)^{-1}.
\end{eqnarray}
The actual observed luminosity density, based on the magnitude $m_{AB}(1600\AA )=24.84$ falls short by a factor 70 of achieving this ionizing flux at the (projected) position of the filament,
even if all the escape fractions take on their most optimistic values. Similar conclusions are obtained from the starburst99 models of object A, where the instantaneous, $4\times10^6$ yr old 
starburst falls short of producing the required yield of ionizations by a factor 67. The continuous star formation model with 14 $M\odot$yr$^{-1}$ produces more ionizing photons 
but still falls short by a factor 22. Stellar winds may be able to increase the escape of ionizing photons from stellar atmospheres by an order of magnitude (e.g., Najarro et al 1996),
but that may still not be sufficient. 

\bigskip

The possibility that object A (or any of the other galaxies) could harbor an AGN that would irradiate the filament with ionizing photons cannot be ruled out. Fluorescence induced by an AGN may cause very large equivalent widths over distances much larger than considered here  (e.g., Cantalupo, Lilly, \& Haehnelt 2012). As mentioned above, the spectral line 
width of the Ly$\alpha$ line  at FWHM=870 kms$^{1-}$
is somewhat large for a standard Ly$\alpha$ emitter, but the broad emission is spatially extended, and so does not in itself suggest that the galaxy hosts an AGN. However, the non-detection of this object in the Chandra x ray data puts limits on the nature
of the underlying source. Luo et al (2010) report a detection threshold for the soft X-ray band ([0.5-2.0] keV) of $1.9\times 10^{-17}$ erg cm$^{-2}$s$^{-1}$. Extrapolating the flux in that band
to the total flux of hydrogen ionizing photons below the Lyman limit would require a very steep slope of $f_{\nu}\propto\nu^{-\alpha}$ with $\alpha\sim 3.3$, in order to produce $4\times10^{55}$ s$^{-1}$ of ionizing photons. The AGN could be compton-thick and emit no x-rays along the line of sight to the observer,
while still irradiating the filament. For an AGN with intrinsic luminosity density (in units of erg s$^{-1}$Hz$^{-1}$) of $L_\nu \propto \nu^{-0.5}$ for
$\lambda < 1050\AA $, and
$L_\nu \propto \nu^{0.5}$ for
$\lambda > 1050\AA $ (following Bolton \& Haehnelt 2007),  capable of inducing the observed Ly$\alpha$ flux in the filament through an optically thin sight-line, 
the unextinguished apparent AB magnitude in the ACS F435W B band would be 23.8, brighter by about 1.5 magnitudes than the observed B-band flux of galaxy A.
Given that only about 1\% of all Lyman break galaxies according to the standard definition are QSOs, one would expect to find 
a QSO on our long slit with a Poissonian probability of about 4\%. We have, however, already another QSO on the slit (to be published in a future paper). The probability for finding  two or more QSOs in the same volume is 0.15\%. 

\subsubsection[]{Shocks and emission from wind-shells}

Shocks, and the emission from wind shells (e.g., Taniguchi \& Shioya 2000;  Mori et al 2004) are further candidates for high equivalent width emission from narrow structures
and could provide Ly$\alpha$ emission through either collisional or photo-ionization channels. 
Paper I  considered the possibility of shocks ionizing the neutral hydrogen in the z=3.34 emitter, but found them unlikely to to explain the observed Ly$\alpha$ emission, as a rather
large shocked area and high velocities were required.
A similar situation exists here, with the Ly$\alpha$ emitting region of the filament having  a much smaller area. We have no observations of shock diagnostics (e.g., the HeII 1640
line) available, nor does the filamentary geometry  suggest a shell or shocked sheet of gas. In the starburst models discussed above ($4\times10^6$ yr old instantaneous starburst or continuous star formation), the mechanical energy from O and B and Wolf-Rayet stars (Leitherer et al 1992), if disposed of in the form of Ly$\alpha$ radiation, could give a significant boost to the total Ly$\alpha$ emission from an underlying stellar population, but not a dominant one. 

\subsubsection[]{Emission from cold gas accretion, powered by cooling radiation}

The filament in the present halo shows certain similarities with the z=3.34 filament discovered in the Rauch et al 2011 object. Both terminate in a faint galaxy,
and - if we accept the premise that the flux in the B-band image of the present case is indeed mainly Ly$\alpha$ emission - neither can be linked causally
to any stellar features.
The possibility can again be raised whether this is actually one of the predicted cold stream filaments thought to dominate the accretion process
at early times and for lower mass galaxies. Paper I suggested that the filamentary structure seen near the 3.34 galaxy
is likely to be seen in fluorescence when exposed to ionizing external radiation from the main galaxy (or possibly a hidden AGN therein, or a tidal tail
leaking ionizing photons), but 
such an explanation may be somewhat less likely in the present case, because of the high surface brightness of the filament. Global, extended cooling
radiation from a relatively low mass halo like the present one (see below) may disfavor a signal strong enough to be detected (e.g., Haiman, Spaans \& Quataert 2000, Fardal et al 2001, Dijkstra \& Loeb 2009), but the radiation may be detectable if it comes from very compact regions,
and the partly neutral cores of individual filaments 
may actually be visible at current sensitivity levels. Faucher-Giguere et al (2010), Goerdt et al (2010, 2012) and Rosdahl \& Blaizot (2012) have studied Ly$\alpha$ emission
from cold accretion filaments, with a focus on predicting the emissivity. These studies agree in that the densest regions of the filaments should be partly neutral, and their Ly$\alpha$ emission should be dominated
by collisional excitation cooling. If we take the observed filament as a z=2.63 cylinder of overdense gas with a length of 16.6 kpc  (omitting the region of the apparent galaxy "F" at the tip) and a diameter of 1 kpc, 
the Ly$\alpha$ flux produced by collisional cooling radiation
at the peak emissivity ($10^{-11.2}$ photons cm$^3$s$^{-1}$, near T=20,000 K, e.g., as given by Faucher-Giguere et al 2010) 
can be written as 
\begin{eqnarray}
F_{ly\alpha} =\left(\frac{\dot{E} V}{4\pi D_L^2}\right)= 1.1\times10^{-25}\left(\frac{\epsilon_{ly\alpha}n_H^{-2}}{10^{-11.2}{\rm cm}^3 {\rm s}^{-1}}\right)\times\nonumber \\
\times\left(\frac{n_H}{1.3\times10^{-5}}\right)^2  
{\rm erg\ cm}^{-2}{\rm s}^{-1},
\end{eqnarray}    
where $\dot{E}$ is the emissivity [erg cm$^{-3}$s$^{-1}$], $V$ is the volume of the filament, $D_L$ is the luminosity distance, and $\epsilon_{ly\alpha}/n_H^2$ is the emissivity of Ly$\alpha$ photons, divided by the square of the total hydrogen density [photons cm$^3$ s$^{-1}$] in the nomenclature of
Faucher-Giguere et al 2010. We find that the observed flux from the filament, $4\times10^{-17}$ erg cm$^{-2}$s$^{-1}$, could  be produced by collisionally cooling gas with a total hydrogen density of 0.24 cm$^{-3}$, if a steady supply of potential energy can keep the gas at the optimum
temperature. Thus,
at such densities, which are at the high end of the expected density range for cold streams, cooling radiation may at least in principle be a viable explanation, because of the compactness of the filament.

\subsubsection[]{Intra-halo star-formation, by young and metal-poor stars}

Alternatively, intrinsic star formation in the filament may be able to produce enough ionizing photons to explain the observed Ly$\alpha$ line strength.
Large Ly$\alpha$ equivalent widths on the order of
a couple of hundred \AA\ and beyond have been
reported previously for Ly$\alpha$ emitting galaxies (e.g., Kudritzki et al 2000;  Malhotra \& Rhoads 2002; Saito et al 2006).
To explain the large equivalent
widths through stellar photoionization requires the stellar population to be hot, young, and  metal poor (e.g., Kudritzki et al 2000; Tumlinson \& Shull 2000; Tumlinson, Shull \& Venkatesan 2002, Schaerer 2002), or to exhibit a top-heavy initial stellar mass function (IMF; e.g., Malhotra \& Rhoads 2002).  Under such circumstances, stellar yields of ionizing photons and the conversion of ionizations into Ly$\alpha$ line
photons (through departures from case B) can be strongly enhanced (e.g., Raiter, Schaerer \& Fosbury 2010; Inoue 2011), even much beyond the 150-700 \AA\ range suggested
here. In fact, the presence of a young population ($4\times10^6$ yr) at least for the starburst of galaxy A is consistent with these models. The blue B-V colors observed for some of the galaxies also agree well with the color range expected of very metal-poor, possibly PopIII stars (Jimenez \& Haiman 2006). 
However, a filament of length $L$, if not formed simultaneously as in the case of a compressed shell or caustic, would  be subtended with a relative velocity $v$, e.g., by star-forming regions trailing or forming behind the galaxy F. Such a protracted unfolding of an extended star-forming
filament would produce an age difference between head and tail  of 
\begin{eqnarray}
t=8.1\times10^7\left(\frac{L}{17 \rm kpc}\right)\left(\frac{v}{200{\rm kms}^{-1}}\right)^{-1} yr. 
\end{eqnarray}
This is a factor $\sim20$ longer than the age of the young stars we have postulated to be present
throughout the filament at the time of the observation, so the fact that the filament is visible at similar surface brightness throughout its length is a bit of a puzzle. It is of course possible that the conditions for the continuous formation of stars causing strong Ly$\alpha$ emission
simply persist for the entire  observable lifetime of the filament.  
Another possible explanation may be found
in the finite HI recombination time scale.
Assuming a gas temperature of $2\times10^4$K, a largely ionized medium, and a density of the emitting filament of 180 times the mean density of the universe at the observed
redshift (this is equivalent to assuming that the halo has collapsed recently),
the recombination time scale of the gas in the filament
\begin{eqnarray}
\tau_{\rm rec}=(N_e \alpha_{rec})^{-1} \sim 5.4\times10^7 {\rm yr} \left(\frac{1+\delta}{180}\right)
\end{eqnarray}
can be of a similar order of magnitude as the likely age of the filament, so even a wave of short-lived star formation traveling with the tip of the tidal tail would lead to a extended period of recombinations and visibility of Lyman $\alpha$ emission. If a short instantaneous
starburst is to blame for the photoionization, then the combination of a finite recombination time scale with the rapid decline in the stellar broad band light would assure that the observed equivalent width would be very large,
similarly large as expected if produced by the more exotic star formation scenarios considered here (see Appendix B).

The dynamical origin of the filament T2 may be understood in analogy with low redshift observations of interacting galaxies. The presence of an apparently tidal stellar feature (T1) near object A may suggest that
the filament is a tidal tail, too. Object F could be the remnant of one of the interacting objects, or a dwarf galaxy that
has formed  at the tip of tidal tail, as has been proposed for low redshift interacting galaxies (e.g., Schweizer 1978, Mirabel, Dottori \& Lutz 1992).  Young stellar populations in star forming clumps along low z tidal tails have also been observed and studied in considerable detail  (e.g., Whitmore \& Schweizer 1995; Smith et al 2008; Karl et al 2010).

In an intriguing alternative scenario, the luminous tail may be gas stripped from object F by ram-pressure when passing through the intergalactic medium, and the increased turbulence may have led to in-situ star formation.
Such a phenomenon may have been observed in the low-redshift intracluster medium (e.g., Hester et al 2010; Yoshida et al 2008, 2012) and is consistent with current theoretical ideas (e.g., Kapferer et al 2008; Tonnesen \& Bryan 2010, 2012), although, as far as we know, there are no studies of such an effect at the level of an individual galactic halo, and at high redshift. Intriguingly, a presence of metal-poor stars may directly reflect an inflow of  low metallicity gas from the ambient IGM/halo gas (log Z/Z$_{\odot}\sim -2.8$ at $z\sim 3$; Schaye et al 2003; Simcoe, Sargent \& Rauch 2004, with pockets of 
lower metallicity persisting at the same redshifts; Fumagalli, O'Meara \& Prochaska 2011), favoring the formation of metal-poor stars even at relatively late times. 

And finally, the astrophysical scenarios just described, namely, the presence  of low metallicity, hot, young stars in the low opacity environment of the
gaseous halo, would lead to an enhanced production and escape of ionizing radiation, favorable properties for the (so far hypothetical) galactic sources required for the  reionization of  hydrogen. 

\section[]{Overall nature of the underlying halo}

\subsection[]{Halo membership}

Because of the faintness of the galaxies, membership of the individual galaxies in the halo hosting the Ly$\alpha$ emission cannot be established beyond doubt. 
We have found above, however, that its Ly$\alpha$ forest decrement indicates that A is at the same redshift as the Ly$\alpha$ emission. Object B, while not as blue as A,
coincides in projection with the end of a Ly$\alpha$ "finger" and thus may or may not belong to the group as well. Object C is very close to A in projection and similar to A in colors, and may be also be responsible for some of the substructure seen in the Ly$\alpha$ emission. Object F looks like a halo member because it appears physically connected to the filament T2, the emission of which, we have argued, consists mainly of Ly$\alpha$ radiation
at the same redshift as the main Ly$\alpha$ line. Object E has similar colors and a similar deviant photometric redshift as A, so it is possibly also
belonging to the group. Keeping in mind the considerable scatter in the relation between stellar and
total halo masses, we may identify A with the main halo galaxy and try to gain a crude idea of the halo mass associated with the stellar mass of A.
Taking the $1.1\times10^9 M_{\odot}$ inferred from the rest frame optical colors using MAGPHYS (see above) as the stellar mass (this ignores the recent
starburst), we use the relations between stellar and halo mass derived by Behroozi et al 2010
(extrapolating to z=2.63) and arrive at a total halo mass of $10^{11.6}M_{\odot}$. We cannot determine the stellar masses for the faint satellites because of a lack
of data in the IR (except for A), but we assume very crudely that the factor $\sim10$ difference between the z-band fluxes of the brightest object A and the faintest object F reflects a decade in halo mass ratios. Then a group of 4-5 members, as suggested by the observations in a $\sim 10^{12}M_{\odot}$ halo, is not unexpected (e.g., Kravtsov et al 2004).

\subsection[]{Faint, extended emitters and Ly$\alpha$ blobs}

As for the distinction between our asymmetric halos and those hosting the bright extended emitters known as Lyman $\alpha$ blobs, it may be primarily one of mass, with our emitters being single, protogalactic  halos, and the blobs being more akin to the progenitors of galaxy clusters (e.g., Yang et al 2010). In the more massive halos, the larger number of  subhalos and the  wider variety  of sources capable of producing  Lyman $\alpha$ emission renders the nature of Ly$\alpha$ blobs more diverse and adds considerable complexity (e.g., Yang et al 2011; Prescott et al 2012, Bridge et al 2012). The detection  of HeII 1640 emission in these objects
(e.g., Scarlata et al 2009; Prescott et al 1012) also strongly points to some non-standard stellar or non-stellar ionization mechanism, and the finding of a diffuse stellar component (Prescott et al 2012) may just be one of the consequences of the multiple interactions we appear to be seeing individually in the smaller halos.

\subsection[]{Faint, extended emitters and more typical Lyman $\alpha$ emitting galaxies}

\begin{figure*}
\includegraphics[scale=0.6,angle=0,keepaspectratio = true]{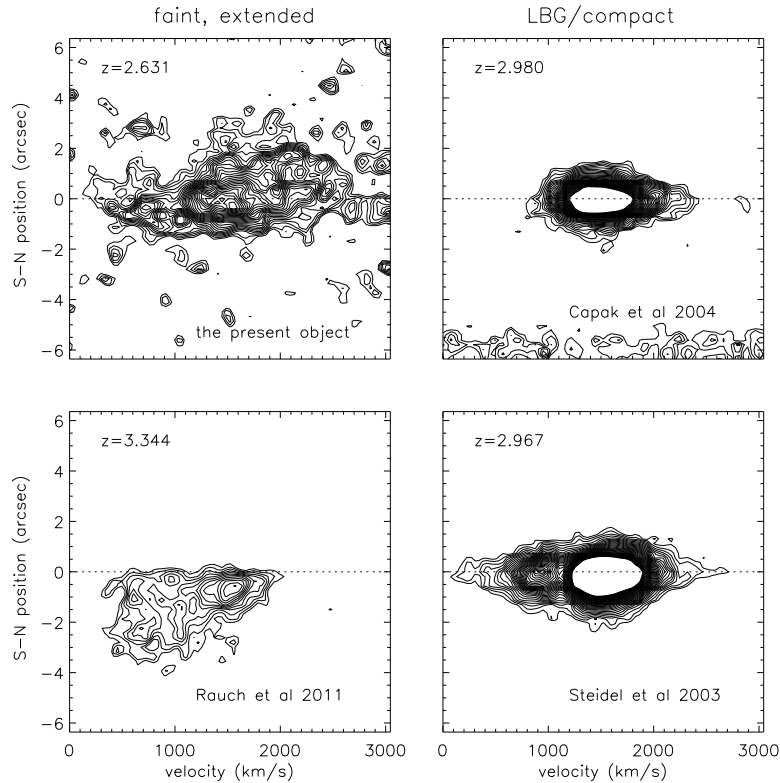}
\caption{Comparison between the two extended emitters (the z=2.63 object at the top left; the Rauch et al 2011 z=3.34 object on the bottom left) and two galaxies that qualify  as "Lyman break" galaxies with Ly$\alpha$ emission,
from a Keck LRIS survey in the Hubble Deep Field North and its immediate vicinity (Rauch et al, in prep.).  
The spectra have been resampled  to match the dispersion and spatial scale, with the
LDSS3 HUDF spectra (left panels) smoothed in both directions to match the coarser pixel resolution
of the binned LRIS HDFN data (right panels). The spectra were put on the same flux contour scale, with
the lowest and highest flux density contours corresponding to $2.1\times10^{-20}$
and $2.4\times10^{-19}$ erg cm$^{-2}$ s$^{-1} \AA$. 
The dotted lines give the approximate position of the peak of the continuum traces. The images are centered in the dispersion direction on the peak emission,
as the systemic redshifts are unknown. 
The spectra of the compact objects on the RHS show the usual red peak with a a sharp drop on the blue side and an extended red shoulder.
Absorption troughs in the continuum just blueward are present but not visible in this contour plot. The bottom system has a (partly suppressed) blue peak as well.
\label{compares}}
\end{figure*}

\begin{figure*}
\includegraphics[scale=.45,angle=0,keepaspectratio = true]{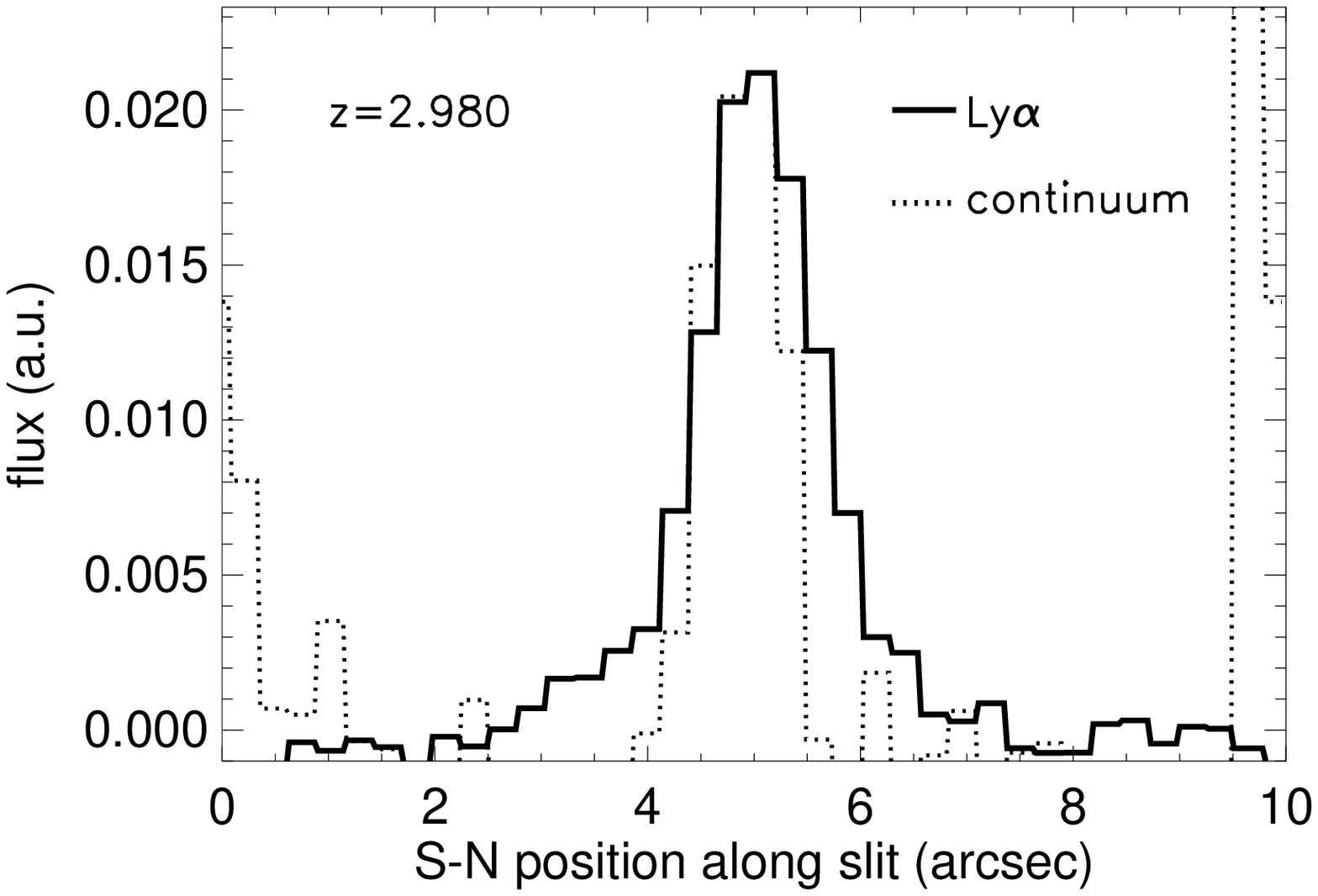}
\includegraphics[scale=.45,angle=0,keepaspectratio = true]{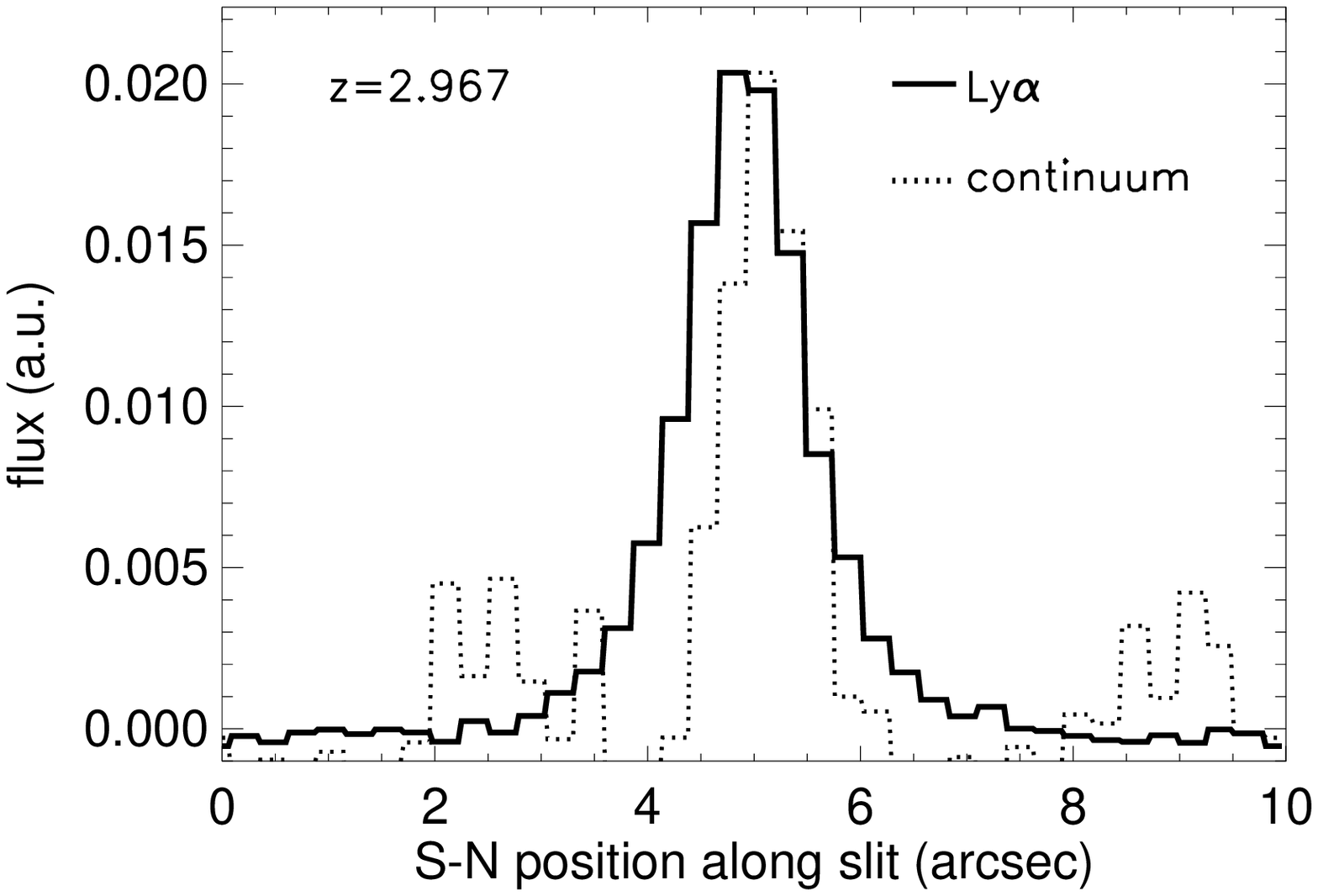}
\caption{Spatial flux profile along the slit direction,  of the two compact Lyman break galaxies J123647.09+620935.8 (left) and J123646.94+621226.1 (right) from fig. \ref{compares}. The Ly$\alpha$ spectral profiles (solid lines) were added up
along the dispersion direction, and are shown after the continuum as measured immediately to the red of the Ly$\alpha$ emission line (dotted line) has been subtracted. The Ly$\alpha$ line
and the continuum are both scaled to have the same peak flux to facilitate a comparison between their respective spatial profiles. While the emission in both is strongly
peaked, the Ly$\alpha$ emission has significantly broader wings that can be traced out to about 2" at this sensitivity level. There is also some asymmetry in that
the Lyman $\alpha$ and continuum profiles do not line up precisely but are displaced spatially along the slit by shifts on the order of a pixel (0.27").
\label{2profiles}}
\end{figure*}

It is instructive to look at the differences in the appearance of the Ly$\alpha$ emission line and other parameters between the extended Ly$\alpha$ emitters
found by paper I and in the present study,  and those "standard" Ly$\alpha$ emitters that dominate typical Lyman $\alpha$ emitter samples. The latter, discovered by broad band color-selection or narrow
band imaging, generally are more highly peaked in the spatial
direction. For comparison, we show, in fig. \ref{compares} our two extended emitters (left panels) and two bright, compact Lyman $\alpha$ emitters (right panels)
that happened
to be accidentally intersected by the long slit during the Keck LRIS blind survey in the Hubble Deep Field North (Rauch et al, in prep.). Color- and luminosity-wise, the latter two
(J123647.09+620935.8 (Capak et al 2004) at z=2.9804, top right panel, and J123646.94+621226.1  at z=2.96707, a.k.a. HDF0D10, (Steidel et al 2003); bottom right panel) can be classified as Lyman break galaxies with Ly$\alpha$ emission. The objects are presented on the same spatial, velocity and flux scale so that a direct comparison can be made. In addition, fig.\ref{2profiles} shows the two profiles from the right column of fig. \ref{compares}, 
collapsed along the dispersion direction. These profiles correspond to the projections of 
the central parts of the surface brightness profiles that have passed through the slit. The solid profiles
show the Ly$\alpha$ emission after the continuum  (measured immediately to the red of Ly$\alpha$) has been subtracted. The continuum profile itself (dotted line), scaled for  comparison to the same peak value, outside of the central arcsecond is clearly less extended than the Ly$\alpha$ emission which can be traced out
to about 2" at the present sensitivity. 
The two objects also show small spatial shifts between Ly$\alpha$ and continuum emission on the order of a (binned) 0.27" wide pixel, or about 2 kpc proper, and the profiles appear mildly asymmetric.
The finding of extended Ly$\alpha$ halos even around these very compact individual galactic sources
is consistent with the individually detected, extended profiles found (for generally fainter) objects in the somewhat deeper 
VLT FORS survey by Rauch et al (2008), and has also been seen in stacks of Lyman break galaxies (Hayashino et al 2004; Steidel et al 2011). 
The difference between the extended, asymmetric profiles on the left of fig.\ref{compares} and the compact, more-or-less symmetric ones of the Lybreak galaxies on the right, may have, as we have argued earlier, several
causes that are likely to be related to the hierarchical nature of the assembly of the halos. Clustering of galaxies will lead to multiple sources
spatially spread out over the extent of the halo. The increased differential motion of the gas stirred by the in-falling galaxies will  open up escape channels in
velocity space  for Ly$\alpha$ line radiation. Galactic collisions may inject stellar tails into the lower opacity
gaseous halo, and, in conjunction with the ambient gas pressure perhaps strip them partly of the ambient interstellar medium. Interactions may
temporarily produce physical holes in the gaseous halos through which Ly$\alpha$ and ionizing radiation may be able to escape. Star formation may be
triggered by merger-induced in-fall and gas-dynamical instabilities in the gaseous halo.  Thus both forms of Ly$\alpha$ emitters may just represent  different stages in the formation of a galactic halo, with the more spatially compact emitters perhaps representing the quiescent phases between interactions, where a single, compact stellar source of ionizing
photons dominates the production of ionizing photons and results in a "compact" source of Ly$\alpha$ emission. 

To test the idea that the central few arcseconds of the compact Ly$\alpha$ emitters are dominated by compact sources
of ionizing radiation, we take advantage of the fact that  the two profiles shown in fig. \ref{2profiles} are among the deepest spatially resolved spectra ever obtained of Ly$\alpha$ emitters, allowing us to compare them individually
to model predictions of the surface brightness profile of Ly$\alpha$ emitters.
The outcome of this comparison, described  in the appendix C, suggests that  single, compact sources of ionizing photons in halos with moderate
expansion velocity (of order 100-200 kms$^{-1}$ in the innermost 10-20 kpc) can formally produce spatial Ly$\alpha$ emission surface brightness profiles
as compact as observed. The Ly$\alpha$ emission profiles examined here do not show signs of the much larger velocity gradients  observed in the absorption troughs of Lyman break galaxies (e.g., Steidel et al 2011), nor do they seem to be consistent with large wind shells as sources of Ly$\alpha$ emission (e.g., Verhamme et al 2008), as these would produce much shallower surface brightness profiles than observed.

\bigskip

\section[]{Conclusions}

Based on an ultra-deep long slit survey of the HUDF and flanking GOODS-S field for faint Ly$\alpha$ emission, we have found a second extended, asymmetric z=2.63 emitter, in addition to the one described in paper I.  The Ly$\alpha$
emitting region coincides spatially with several  objects detected in broad band HST ACS images from the GOODS survey. Multiple photometric
redshifts  for some of the objects from the literature disagree not only among each other but also with the spectroscopically determined
redshift for the Ly$\alpha$ emission. However, the brightest galaxy, A, shows a Ly$\alpha$ forest decrement as expected for the z=2.63 redshift. 
The rest frame UV colors of this object are best fit with a very young starburst which may have compromised previous photometric redshift determinations.
Several other objects have similar B-V colors.
Interactions between the merging progenitors in a common, Milky Way-sized halo may account for several unusual findings, including the sprawling Lyman $\alpha$ emission,
the spatially irregular stellar components, the blue colors and likely presence of young stars in several objects,  and the evidence for a possibly Ly$\alpha$ emitting filamentary structure connected to a faint continuum source.
The 17 kpc long filament (which is mostly outside of the spectrograph slit), and another more diffuse emission region that coincides with a narrow
finger of Ly$\alpha$ emission seen in the 2-dimensional spectrum, are detected by their excess light in the B band, but not in any other broad band.
If the excess radiation is interpreted as Ly$\alpha$ emission, the resulting large inferred equivalent widths either suggest non-stellar processes like Ly$\alpha$ fluorescence,
cooling radiation from a cold accretion filament, or the formation of otherwise undetected hot, metal-poor stars, perhaps in a tidal tail or the wake
of a ram-pressure-stripped galaxy. Unlike the case of the Rauch et al (2011) z=3.34 galaxy, where a gaseous filament falling in
to the main galaxy appeared to fluoresce in response to ionizing radiation escaping from the same galaxy, the non-detection
of sufficiently powerful stellar or AGN sources in the present case disfavors the possibility of fluorescence as the
source of the filamentary Ly$\alpha$. 
Cooling radiation in a dense, cold accretion stream may give an energetically feasible explanation for the filamentary Ly$\alpha$. However,
the linear structure and trailing appearance of the filament resembles the features seen in low redshift
tidal tails, or the formation of stars in a turbulent, stripped wake, trailing a satellite galaxy moving through the main
common gaseous halo.

In the present case, other than through the possible detection of optically thin Ly$\alpha$ emission in the Ly$\alpha$ "finger", we have no direct evidence for the escape of ionizing radiation, which, because of the low redshift and
faintness may be hard to obtain observationally. However, physical circumstances in an interacting halo should be quite favorable for the
escape of ionizing photons. Interaction between halo members may cause disturbances in the gaseous halo and intra-halo star formation in tails and wakes, either of which serves to reduce the opacity for ionizing radiation.
The formation of preferentially hot, metal-poor stars, perhaps fed directly by the low metallicity IGM in the halo, would be 
characterized by a higher yield of ionizing photons, with a harder spectrum.
Interactions, triggering these processes increase rapidly when going toward higher redshift.
All these features go in the same direction, suggesting {\it disturbed
galactic halos may be important contributors of ionizing photons during the epoch of reionization}.

We have briefly discussed the  difference between these extended, asymmetric emitters, and other classes of Ly$\alpha$ emitters. 
Observational samples, irrespective of how they were obtained,  tend  to be dominated by spatially symmetric, compact sources. Examining two
bright Ly$\alpha$ emitting Lyman break galaxies we show that, nevertheless, even those objects have extended Ly$\alpha$ envelopes 
surrounding the continuum sources of ionizing photons. Comparison with simple models shows that the observed Ly$\alpha$ surface brightness profiles can be
modeled by embedding single point sources of ionizing photons in intact spherically symmetric HI halos. Very extended stellar sources are not indicated,
and radiating wind shells appear inconsistent with the profiles, at least on the $\sim 20$ kpc scales that we can observe.
We interpret these "standard" Ly$\alpha$ emitters as halos in the "off" stage, where one galaxy strongly dominates the emission and interactions are not currently important. 
The differences between our faint, extended emitters and  the more luminous, rarer Ly$\alpha$ blobs may mainly depend on mass and energetics, with the
blobs being more massive halos containing a larger variety of more powerful sources of Ly$\alpha$ radiation like AGN, and radiogalaxies. In contrast, the fainter and more numerous objects that we appear to be discovering from a combination of
deep spectroscopy and archival imaging are more likely to belong to the progenitors of normal present day galaxies. 
While considerable uncertainties remain, these observations have the potential of giving us rare views of the actual formation of  such objects.

\section*{Acknowledgments}

We acknowledge helpful discussions with Bob Carswell,  Hsiao-Wen Chen,  Jeff Cooke, Masami Ouchi, and Francois Schweizer.  We  thank the staff of the Las Campanas Observatory and the Keck Observatory for their help with the observations,
and thank Luke Barnes for providing us
the results of his simulations.
 MR is grateful to the IoA in Cambridge and to the Raymond and
Beverley Sackler Distinguished Visitor program for hospitality and support in summer 2011, when some of this work was done. He further acknowledges support from the National Science
Foundation through grant AST-1108815.  GB has been supported by the Kavli Foundation.

\newpage

\smallskip

\pagebreak

\newpage

\appendix

\section[]{Determining the precise position of the slit}

\begin{figure*}
\includegraphics[scale=.45,angle=0,keepaspectratio = true]{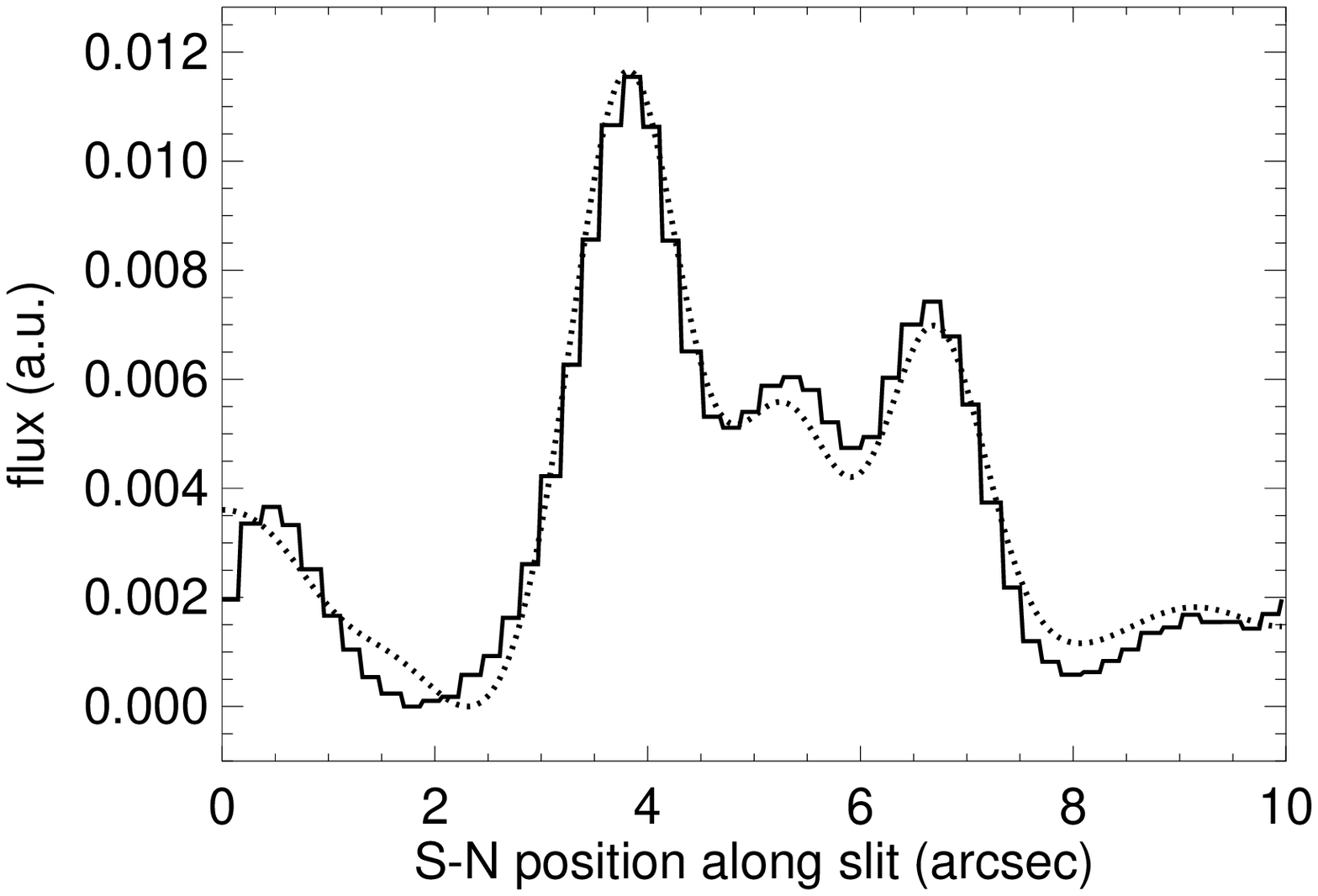}
\includegraphics[scale=.45,angle=0,keepaspectratio = true]{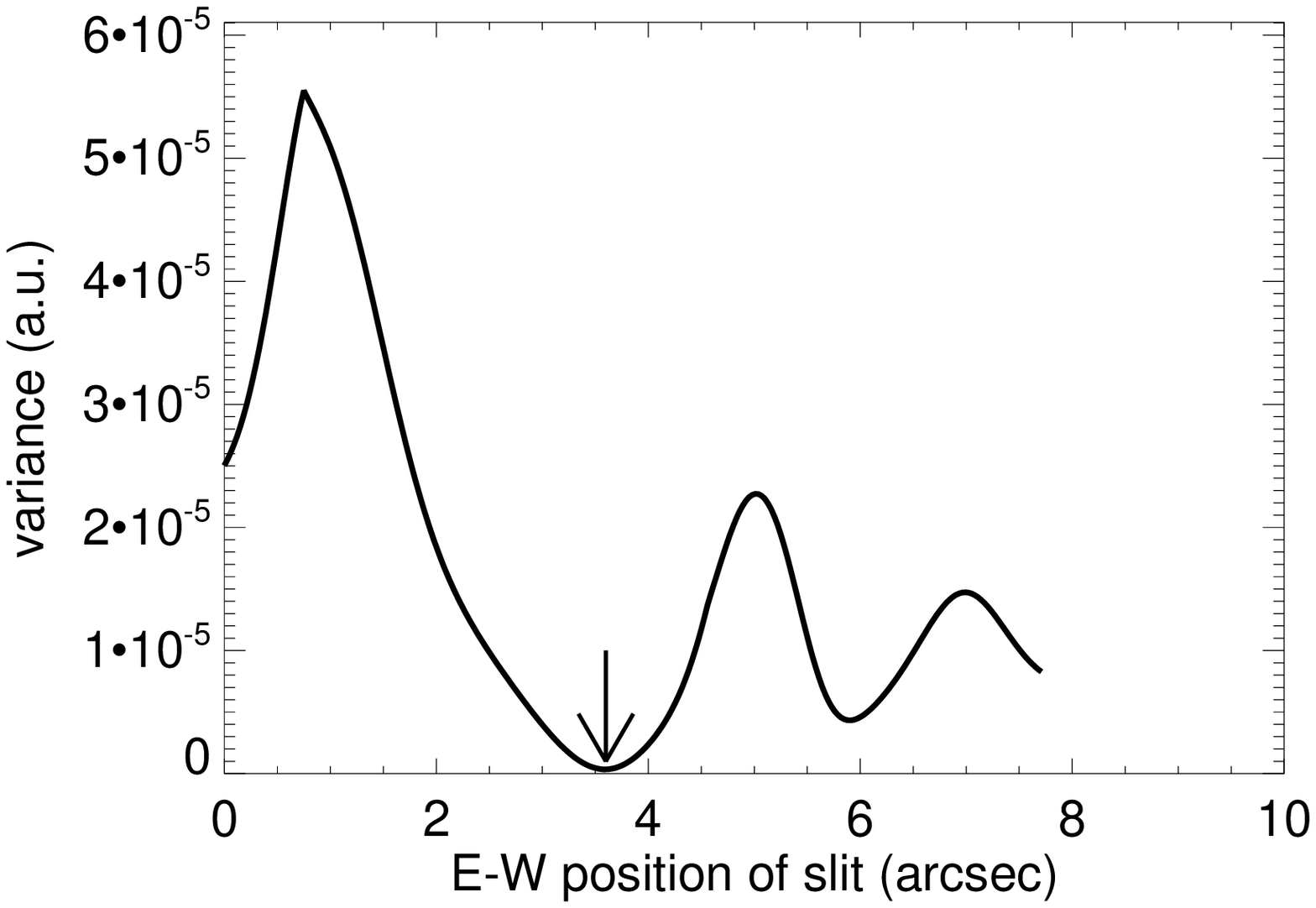}
\caption{LHS Panel: Best match between the flux profile along the slit of the spectrum projected along the dispersion direction (solid histogram), and the flux profile along the slit of all light from the target field transmitted through the slit, projected perpendicular to the slit direction (dotted histogram). A good match was obtained for a Gaussian smoothing
kernel with 1.05" FWHM, an effective slit width of 2.25". The useful spectral range was chosen to be 4800 - 6182 \AA .  RHS panel:  The plot shows how the variance between the two profiles changes as the slit moves
in a E-W direction (i.e. perpendicular to the slit direction) across the sky. The abscissa is the position of the east slit jaw with respect to the edge of image \ref{prettypics} (center and rightmost panels). The minimum (arrow) corresponds to the profiles shown in the LHS panel. The corresponding best fit slit position is shown in fig. \ref{prettypics}.
\label{app1}}
\end{figure*}

Because of the length of the slit used during the observation (8.3"), departures from straightness, uncertainties in the initial slit positioning, guiding errors,
and differential atmospheric refraction, the precise placement of the slit can be uncertain to within a significant fraction of the slit width.
Through-slit-images taken before and after the spectroscopic observation may help, but these usually lack depth, and there is no guarantee that the slit remained at the imaged position during the spectroscopic exposures.
It is possible, however, to determine the actual, effective placement of the slit on the sky to considerable precision after the fact by
comparing projections of deep space based images perpendicular to the direction of the slit with the distribution of light along the slit in the spectrum. In the current case, an HST ACS F606W image
as shown in fig. \ref{prettypics} was smoothed with a Gaussian kernel with variable FWHM values to imitate ground-based seeing and any fluctuations in the position of the slit. Imaginary slits with widths near the nominal slit width (2") were then moved
across the images, summing the fluxes from all the pixels covered by that slit into a one-dimensional distribution of flux along the slit direction. 
The shape of this profile, reflecting any partial or full coverage of galaxies intersected, should mirror the flux along the slit direction of the summed spectrum,
if covering the same range of wavelengths as the broad band image. The goodness of the match will depend on the actual position of the slit, the effective slit width, the correct assumptions about
smoothing. In principle, a grid of these values can be run, and the best model chosen by a multi-dimensional minimization process. In reality, we looked at a range of values by eye,
and determined that the slit position, our main aim, was not strongly dependent on either the smoothing or the slit width. Thus, the best match between image and slit profile was chosen visually
occurred close to a smoothing FWHM of 1.05", and an effective slit width of 2.25". Our spectrum did not cover the full ACS F606W band, but we found that changing the spectral length of the wavelength
region used to compare with the F606W image did not strongly alter the light profile. Going further to to blue, the galaxy "C" however,  turns fainter much faster than the other objects,
suggesting that using the F435W image instead would have required a more thorough match between the spectral and imaging wavelength bands.  Having fixed the other parameters,
the optimal slit position was determined as the minimum of the RMS difference between the normalized fluxes of the projected imaging profile and the spectral profile. The resulting profiles are shown in 
fig. \ref{app1}. The best fit slit position and width is used in figs. \ref{prettypics}.


\newpage

\section[]{The effect of a finite recombination time scale on the observed Ly$\alpha$ flux and equivalent width}

\begin{figure*}
\includegraphics[scale=.45,angle=0,keepaspectratio = true]{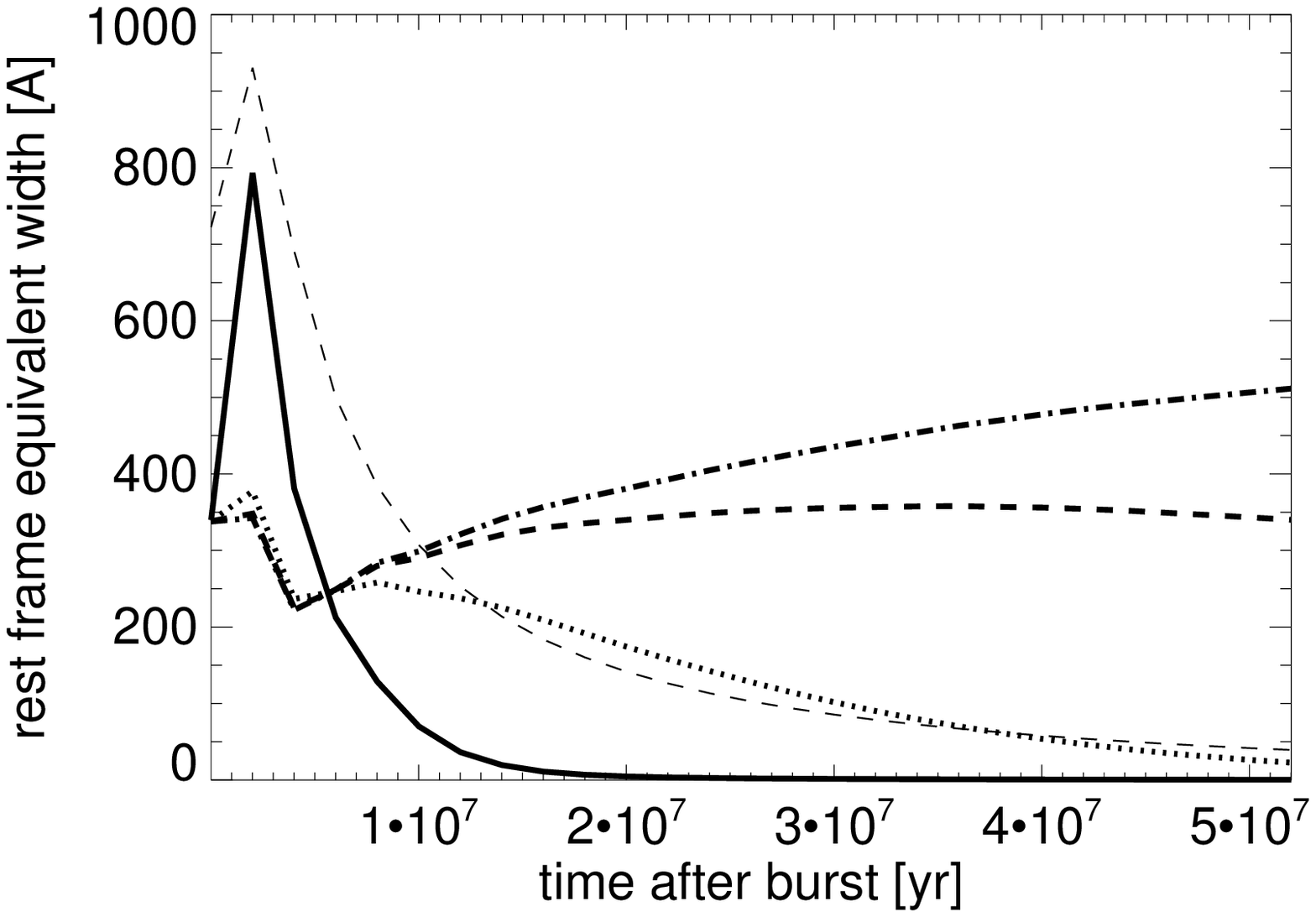}
\caption{. The figure shows the predicted rest frame equivalent width versus time after a starburst, for a z=2.63 instantaneous starburst with $7\times 10^7 M_{\odot}$ in stars, with a metallicity
 of Z=0.0004. The models here assume instantaneous ionization, but recombination (and emission of Ly$\alpha$) with a recombination time scale $\tau_{rec}$. The thick solid, dotted, dashed and dash-dotted graphs
are for $\tau_{rec}$ = $10^6$yr,  $10^7$ yr, $5.4\times10^7$ yr, and $5\times10^8$yr, respectively. Because of the delayed emission of Ly$\alpha$ photons, the equivalent width rises to considerable values, while the flux in the B band drops by a factor 23 over the first $5\times10^7$ yr. The rapid variation during the first $10^7$ yr is cause by the delayed rise of the stellar (B-band) continuum luminosity relative to the production rate of ionizing photons. In addition to the equivalent widths, the plot also  shows the evolution of the Ly$\alpha$ line flux for the $\tau_{rec}=5.4\times10^7$ yr case, as the thin dashed line, in arbitrary units.
 \label{app2}}
\end{figure*}

After the injection of ionizing photons by a short starburst into the surrounding interstellar and intergalactic gas the ionized gas recombines, accompanied by the emission of a proportional fraction of Ly$\alpha$ photons. The recombination process will delay and drag out the emission of Ly$\alpha$ photons, relative
to the production rate of the ionizing photons and the stellar continuum. Assuming that the gas is near the virial overdensity ($1+\delta\sim 180$ for a recently collapsed halo at $z= 2.63$), the recombination time scale is 
\begin{eqnarray}
\tau_{\rm rec}=(N_e \alpha_{rec})^{-1} \sim 5.4\times10^7 {\rm yr} \left(\frac{1+\delta}{180}\right).
\end{eqnarray}
As the timescales for the escape of Ly$\alpha$ photons through scattering even from an optically thick halo
are usually shorter than that (Roy et al 2010; Xu, Wu \& Fang 2011), recombination will dominate the evolution of the emergent Ly$\alpha$ flux. The delayed release of the Ly$\alpha$ photons will 
render the emitter visible for a longer time and may explain, how a wave of instantaneous star-formation passing through a filament (e.g., in a turbulent wake) can maintain similar flux levels along the filament, even though the starburst may be
of much shorter duration than the time it takes for the filament to extend to its current length.
The time evolution of the equivalent width of Ly$\alpha$ (as the ratio between Ly$\alpha$ line flux and 
continuum flux density)  will dependent on both the changes of the (retarded) line flux and the more quickly decaying continuum flux. If the broad band continuum flux fades more rapidly than the recombinations can occur, the oberved
equivalent width can rise and stay high for a significant amount of time. Fig. \ref{app2} shows the predicted rest frame equivalent width versus time after a starburst, for a z=2.63 instantaneous starburst with $7\times 10^7 M_{\odot}$ in stars,  a metallicity of Z=0.0004, and otherwise default assumptions, simulated with the STARBURST99 code. The time evolution of the Ly$\alpha$ flux was simply convolved with an exponential kernel with width $\tau_{rec}$. Even though the B-band continuum luminosity drops by a factor 23 over the first $5\times10^7$ yr, the Ly$\alpha$ flux drops less rapidly, so the equivalent width stays high or remains rising for all but the shortest one of the time scales shown. While the instantaneous equivalent width would be as high as 800\AA , the recombination delay would make this stage unobservable, and the rapid decay would render finding a galaxy in this state quite rare.
The plot shows the effect of different densities, with the dashed line marking the recombination time for an overdensity of 180 at z=2.63, and the longer (shorter) timescales corresponding to lower (higher) gas densities.
The effect may explain how Ly$\alpha$ radiation, after a starburst, may remain visible longer than the stellar continuum light. It may also account for some of the observations of unusually large equivalent widths of Ly$\alpha$ emitters
(e.g., Malhotra \& Rhoads 2002), without having to invoke exotic stellar populations or the peculiar propagation of Ly$\alpha$ photons in the presence of a clumpy and dusty medium (e.g., Neufeld 1991, Hansen \& Oh 2006).

\newpage

\section[]{Constraints from surface brightness profiles on the emission mechanism of  Ly$\alpha$ emitters }

Most Ly$\alpha$ emitters, regardless of luminosity or how they are detected, and for the entire range of luminosities observed, have  singly-peaked, somewhat extended  surface brightness profiles (e.g., Rauch et al (2008). In the spectral direction  they show the well-known double-humped emission profiles with a dominant red peak, showing a broad shoulder to the red and 
a sharp drop in the blue (e.g., Tapken et al 2007; Rauch et al 2008, (Kulas et al 2011, Yamada et al 2012). A hint of the blue peak can often be discerned (see e.g., fig. 3 of Rauch et al 2008), but at greatly reduced flux levels compared to the red peak. Profiles where the blue peak dominates over the red one are  rare, and are generally thought to signify in-falling gas
(for an example, see Rauch et al 2008, fig. 18).

The fluxes and 
likely masses of the typical Ly$\alpha$ emitters and their spatial compactness suggest that they are consistent with being powered by stellar ionization  from a single internal source,
at least in the innermost $10 - 20$ kpc. However, because of a lack of deep 2-d spectroscopy this assumption has
not been directly tested so far, and direct comparisons between individual observed and simulated surface brightness profiles
have not been performed.

Simplified, spherically symmetric models have explored several possible basic scenarios for the formation of the Ly$\alpha$ emission line. The models examined fall under the basic categories of outflows and
infall of gas. 
Among the outflow models  the  emission of Ly$\alpha$ from wind shells (e.g.,Pettini et al. 2000, 2002; Verhamme, Schaerer, \& Maselli
2006; Schaerer \& Verhamme 2008; Verhamme et al. 2008; Quider et al. 2009) has received the most attention. These models are capable of giving quite realistic representations of the one-dimensional spectra
of Ly$\alpha$ break galaxies (e.g., Tapken et al 2007, Verhamme et al 2006, 2008). However, Barnes \& Haehnelt (2010) have pointed out that shells generally lead to a shallow surface brightness profile, which may be different from what is 
observed in actual 2-d observations of Ly$\alpha$ emitters (e.g., Rauch et al 2008). Collapsing protogalactic halos have been modeled by Dijkstra et al 2006 and Barnes \& Haehnelt 2009, 2010).  These model do produce peaked emission
with extended low surface brightness aprons, but, at least in the absence of intergalactic absorption,  they do not get the correct preponderance of the red peak over the weaker (or absent blue peak). However, as the spectral features of those spherically symmetric models are reversible
under a sign change of the velocity profile, these models can also be used to represent outflowing motion.
\medskip

Because of the considerable depth of our spectra (fig. \ref{compares}), we can test the various hypotheses about the origin of Ly$\alpha$ directly on individual galaxies without having to
resort to stacks. 

\begin{figure*}
\includegraphics[scale=.45,angle=0,keepaspectratio = true]{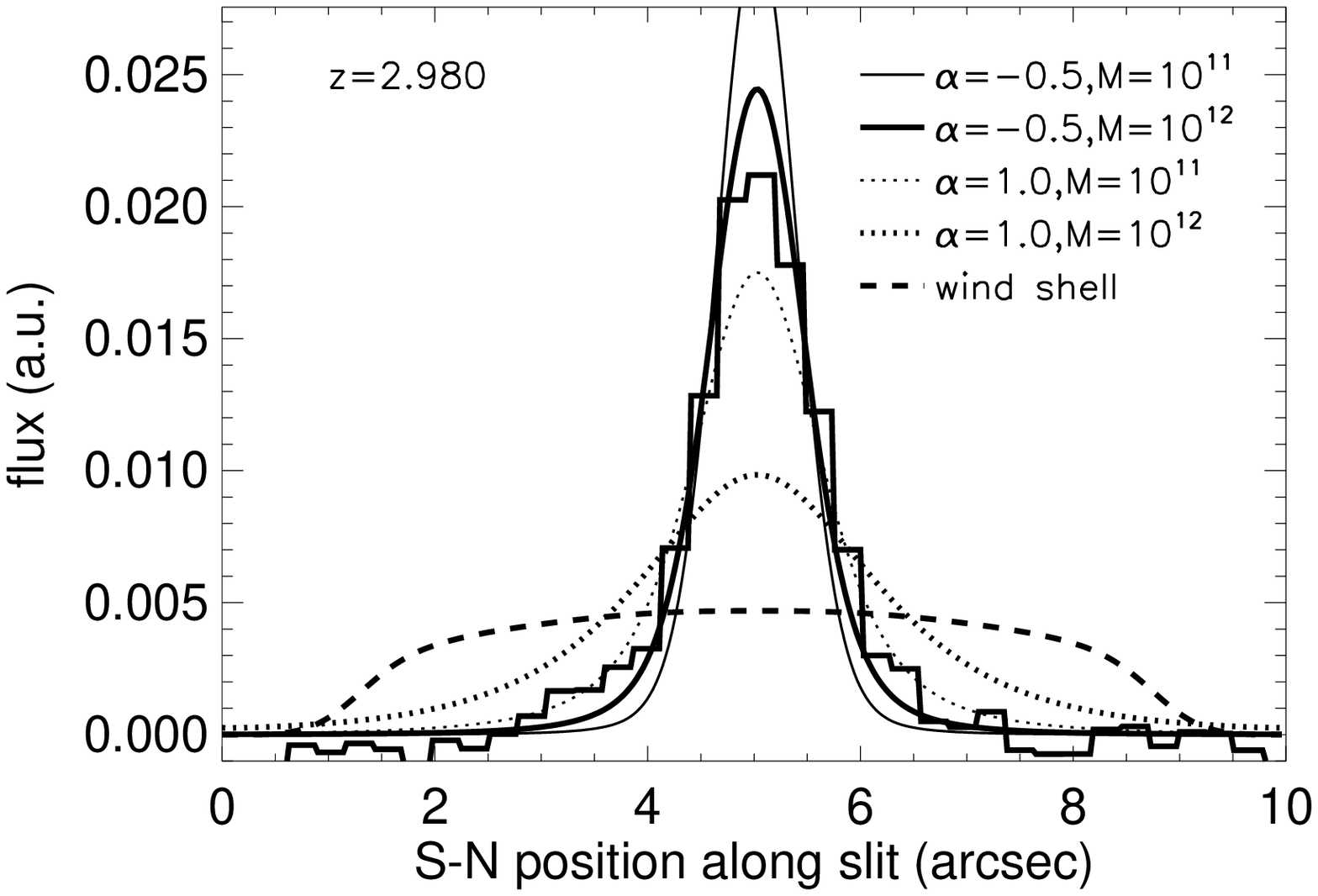}
\includegraphics[scale=.45,angle=0,keepaspectratio = true]{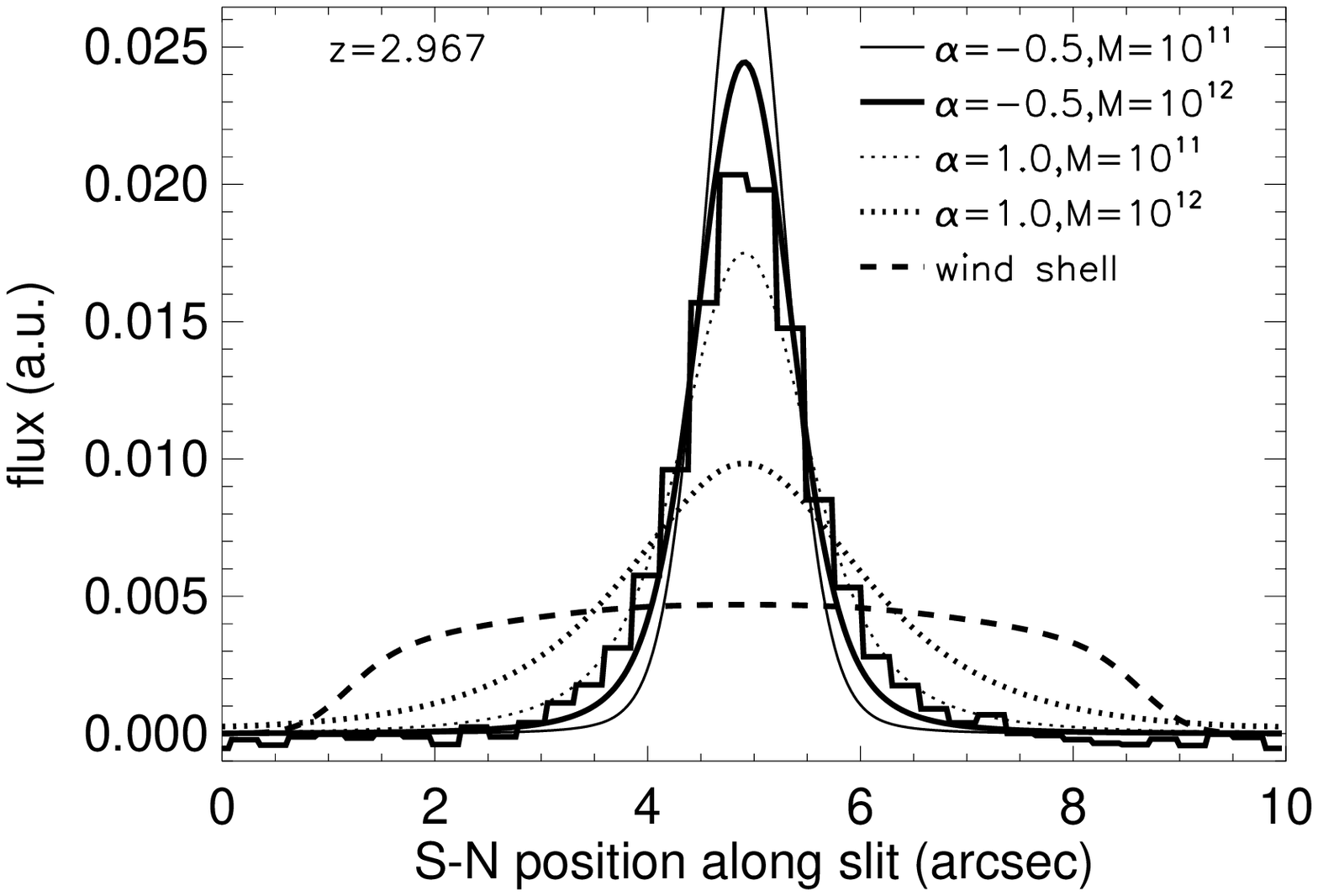}
\caption{The same two spatial Ly$\alpha$ profiles along the slit as in fig. \ref{2profiles} (solid histogram), but now overplotting  the predicted profiles expected from
the surface brightness distributions of various models by Barnes \& Haehnelt 2010. These models assume gaseous halos in different
kinematic stages, surrounding a point source of ionizing radiation. The thin and thick solid lines
show the $\alpha$=-0.5 models for total halo masses $10^{11}$ and $10^{12} M_{\odot}$, respectively. For the observed red-dominant peaks this model corresponds to a decelerated outflow. The thin and thick dotted lines are for the $\alpha$=1 models
with halo masses $10^{11}$ and $10^{12} M_{\odot}$ (the wider profile is from the higher mass halo) and correspond to accelerating outflows. The dashed line is the N(HI)=$2\times10^{22}$ cm $^{-2}$ wind shell of Barnes \& Haehnelt (2010)'s fig. 6. The profiles
have been smoothed with a FWHM=0.9" wide kernel.
\label{bh2profiles}}
\end{figure*}

Fig.\ref{bh2profiles} again shows the Ly$\alpha$ emission profiles from fig. \ref{2profiles}, now with several different slit-profiles from models by Barnes \& Haehnelt (2010) overplotted.  The models are fits only in so far as the overall flux has been
normalized - i.e., they all have the same area under the curves, and their spatial position along the slit has been slightly adjusted by hand to account for the fact that the observed Ly$\alpha$ profiles are not quite lining up with the
origin defined by the continuum traces.
The Barnes \& Haehnelt inflow models introduced by Dijkstra et al 2006, are parametrized by a power law dependence of the radial velocity 
on the radius, i.e.,  $v \propto - r^{\alpha}$,
with power law indices $\alpha$ = -0.5 for accelerated infall of massless shells and $\alpha$=1.0  for spherical top-hat collapse. As pointed out above, only a few emitters show dominant blue components
of the double-humped profile (Rauch et al 2008; Kulas et al 2012; Yamada et al 2012). There are several explanations for this, the relative importance of which is not yet fully
understood. 
Intergalactic absorption would erode blue peaks ( Dijkstra, Lidz  \& Wyithe 2007; Laursen, Sommer-Larsen \& Razoumov 2011), as would an outflow of gas away from the stellar source of ionizing photons (e.g., Tenorio-Tagle et al 1999; Kunth et al 1998; Verhamme et al 2006).
In the case of a generic outflow scenario,  reversing the velocity vector will flip the dominant blue peak from the infall picture into a dominant red one, but the surface brightness and column density profiles remain unchanged. We can thus use the Barnes \& Haehnelt models
to study outflow situations as well, keeping in mind that the $\alpha$=-0.5 surface brightness profiles now correspond to a decelerating outflow (or decelerating expanding halo), and the $\alpha=1.0$ case becomes an accelerating
outflow (expansion). The resulting profiles as seen through the spectrograph slit and collapsed perpendicular to the slit like the real data
are shown in fig. \ref{bh2profiles}.
The profiles
have been smoothed with a FWHM=0.9" wide kernel to account for the finite observed size of the stellar continuum (1.0" and 0.9" for the two galaxy continua just redward of Ly$\alpha$ emission line). 

The two smooth solid lines
show the $\alpha$=-0.5 models (decelerated outflow) for total halo masses $10^{11}$  and $10^{12} M_{\odot}$. After the global scaling with flux the lower mass halo seems to have a surface brightness profile too concentrated to match the observed profile 
within the central few arcseconds that we can address here.  Better agreement (although still with slightly too narrow wings)
is obtained from the $10^{12} M_{\odot}$ model. The dotted lines are for the $\alpha$=1 (accelerated outflow) models
with the same halo masses $10^{11}$ and $10^{12} M_{\odot}$ (the wider profile is from the higher mass halo). For those (accelerating) models, the lower mass one (thin dotted line) gives a fit approximately as good as the decelerating, $\alpha$=-0.5, higher mass case, but now overestimating the width of the observed surface brightness profiles somewhat.
The dashed line is the N(HI)=$2\times10^{22}$ cm $^{-2}$ wind shell of Barnes \& Haehnelt 2010's fig. 6. 

For the $\alpha=-0.5$ with their very strongly peaked narrow profiles, the agreement with the observation is obviously sensitive to the degree of smoothing involved. We found that a slightly larger Gaussian smoothing window (1.5" instead of the
1" FWHM) gives almost perfect agreement with the observed profile, but we did not see such a smoothing physically warranted to implement it. In reality, however, the source
of ionizing photons would be a galaxy subject to patchy emission and peculiar motions which soften the very sharp emission peak that characterizes the {\it decelerated} models. 
Given the number of additional degrees of freedom, including the column density, concentration parameter, and the possibility of other, more
physically motivated velocity-radius relations, the current snapshot of model parameters cannot be expect to deliver any strong constraints on the nature of the velocity field. The relatively concentrated spatial surface brightness profiles suggest, however, that the overall velocity gradients
involved in the formation of the Ly$\alpha$ line do not exceed a couple hundred kms$^{-1}$, which is also indicated
by the small velocity offsets observed between the Ly$\alpha$ emission line and the systemic velocity for bright, compact
Ly$\alpha$ emitters (e.g., Hashimoto et al 2012).
Large velocity gradients of up to 800 km s$^{-1}$ as seen in low ionization absorption troughs of Lyman break galaxies (Steidel et al 2011) do not seem to contribute noticeably to the Ly$\alpha$
emission line formation (see also Dijkstra \& Hultman Kramer 2012). The wind-shell model, at least in its implementation as a large scale shell with large covering factor, while producing a
realistic one-dimensional spectral line profile, appears inconsistent with the observations on account of its overly extended predicted surface brightness profile. 

We conclude that, irrespective of the precise velocity field,  the inner $\sim 20$ kpc of the emission surface profiles of strong Ly$\alpha$ emitting galaxies can be  modeled by point sources of ionizing radiation embedded in slowly expanding,
optically thick gaseous halos.

\bsp

\label{lastpage}

\end{document}